\documentclass{aa}

\usepackage{graphicx}
\usepackage{txfonts}
\usepackage[breaklinks]{hyperref}

\hypersetup{
    colorlinks=true,      
    linkcolor=red,        
    citecolor=blue,       
    filecolor=magenta,    
    urlcolor=cyan,   
	pdftitle={Synthetic observations of internal gravity waves in the solar atmosphere},
	pdfauthor={Vigeesh}, 
	pdfsubject={Astronomy and Astrophysics - The Sun},   
	pdfkeywords={Magnetohydrodynamics (MHD), Sun: atmosphere, Sun: chromosphere, Sun: granulation, Sun: magnetic fields, Sun: photosphere, waves, internal gravity waves}
}

\usepackage{ulem}
\titlerunning{Synthetic observations of internal gravity waves in the
        solar atmosphere}
\authorrunning{Vigeesh \& Roth}

\usepackage{color}

\begin{document} 

\title{Synthetic observations of internal gravity waves in the solar
        atmosphere}

\author{G. Vigeesh
          \inst{1}
          \and
        M. Roth\inst{1}
       }

\institute{Leibniz-Institut f\"{u}r Sonnenphysik (KIS),
        Sch\"{o}neckstrasse 6, 79104 Freiburg, Germany\\
           \email{vigeesh@leibniz-kis.de}
                }

   \date{Received ; accepted }


\abstract 
{} 
{
We study the properties of internal gravity waves (IGWs) detected in 
synthetic observations that are obtained from realistic numerical 
simulation of the solar atmosphere.
} 
{
We used four different simulations of the solar magneto-convection
performed using the {CO$^{\rm 5}$BOLD} code. A magnetic-field-free model and
three magnetic models  were simulated. The latter three models start with an initial vertical, homogeneous
field of 10, 50, and 100 G magnetic flux density, representing different
regions of the quiet solar surface. We used the NICOLE code
to compute synthetic spectral maps from all the simulated models for the
two magnetically insensitive neutral iron lines
\ion{Fe}{I}~$\lambda\lambda$~5434~\AA~and~5576~\AA. We carried out Fourier
analyses of the intensity and Doppler velocities to derive the power,
phase, and coherence in the $k_{h}-\omega$ diagnostic diagram to study
the properties of internal gravity waves.
}
{
We find the signatures of the internal gravity waves in the synthetic
spectra to be consistent with observations of the real Sun. The effect of
magnetic field on the wave spectra is not as clearly discernible in
synthetic observations as in the case of numerical simulations. The
phase differences obtained using the spectral lines are significantly
different from the phase differences in the simulation. The phase
coherency between two atmospheric layers in the gravity wave regime is
height dependent and is seen to decrease with the travel distance
between the observed layers. In the studied models, the lower
atmosphere shows a phase coherency above the significance level for a
height separation of $\sim$400~km, while in the chromospheric layers it
reduces to $\sim$100--200~km depending on the average magnetic flux
density.
} 
{
We conclude that the energy flux of IGWs determined from the phase
difference analysis may be overestimated by an order of magnitude.
Spectral lines that are weak and less temperature sensitive may be
better suited to detecting internal waves and accurately determining
their energy flux in the solar atmosphere.
}

\keywords{Magnetohydrodynamics (MHD) -- 
                 Sun: atmosphere -- 
                         Sun: chromosphere -- 
             Sun: photosphere --
             Sun: magnetic fields --
             Waves
               }

\maketitle
   
\defcitealias{2017ApJ...835..148V}{Paper I}
\defcitealias{2019ApJ...872..166V}{Paper II}
%

\section{Introduction}\label{s:introduction}
Several observations have provided strong evidence for the presence of
internal gravity waves (IGWs) in the solar atmosphere
\citep{2008ApJ...681L.125S,
           2011A&A...532A.111K,
           2014SoPh..289.3457N}.
These waves owe their existence to the proximity of an unstable layer adjacent
to a stable environment. Penetrative flows from the unstable convective
layer overshooting into the stable photosphere that lies above excite IGWs,
among other magneto-atmospheric waves, and are thought to propagate
higher up in the atmosphere.

In the low solar chromosphere, especially in the internetwork region, IGWs are thought to contribute to the ultraviolet background at
spatial scales corresponding to meso-granulation
\citep{1997A&A...324..704S, 
           2003A&A...407..735R}. 
\cite{2008ApJ...681L.125S} 
reported that internal waves are found to be locally suppressed in
strong magnetic field regions, while being abundantly present in magnetically
quiet regions of the solar atmosphere. Despite providing a significant
fraction of wave energy flux in the photosphere, the existence and role
of IGWs in the upper layers of the atmosphere in the presence of magnetic
field is still unclear.
\cite{2010MNRAS.402..386N,
          2011MNRAS.417.1162N} 
showed that the waves couple to other magneto-atmospheric waves.

More recent work by 
\citet[hereafter Papers I and II]
      {2017ApJ...835..148V,
          2019ApJ...872..166V} 
looked at IGWs generated in the solar atmosphere and investigated the
effect of magnetic fields on their propagation. These latter authors showed that the
emergent IGW spectra in the near-surface photospheric layer is
unaffected by the presence or strength of the magnetic field.
Internal gravity waves are generated with a considerable amount of wave flux independent of
the average value of the magnetic flux density in the near-surface
region of the quiet Sun, where the waves likely originate. However,
consistent with the observations, the propagation in the upper layers is
influenced by the magnetic fields present there. It was shown that the
subsequent coupling to Alfv\'{e}nic waves is unlikely in a magnetic
environment permeated with predominantly vertical fields, like in the
network regions
\citep[see][for a recent review]{2017SSRv..210..275B},
and therefore they may not directly or indirectly contribute to the
heating of low plasma-$\beta$  layers (where the ratio of gas to magnetic pressure is less
than unity) in the network-like regions of the solar atmosphere.

As their propagation properties are dependent on the average magnetic
field in the higher layers, IGWs could be used to measure the average magnetic field properties of the upper solar
atmosphere. Consequently, they may be used for monitoring the long-term
evolution of the average magnetic field of the  quiet Sun independently of
direct magnetic field measurements. In this paper, we look more closely
into the details of how these waves can be observed and discuss
the implications of using them for diagnostic purposes. We use realistic
numerical simulations of the solar atmosphere to compute the emerging
spectra in the two iron lines {\ion{Fe}{I}}~$\lambda\lambda$~5434~\AA{}
and 5576~\AA{}. Furthermore, we investigate the acoustic-gravity wave spectrum and compare it directly with real Sun observations.

The paper is organised as follows. In Sect.~\ref{s:numerical_setup}, we
discuss the numerical models that we use. In
Sect.~\ref{s:synthetic_spectra}, we describe the synthetic
observables that are derived from the numerical models. A short
description of the analysis is presented in Sect.~\ref{s:analysis}. In
Sect.~\ref{s:results}, we report the results and discussion. The
conclusion of the work is provided in  Sect.~\ref{s:conclusion}.

\section{Numerical simulations}\label{s:numerical_setup}
We carry out full forward modelling of near-surface solar
magneto-convection using the {CO$^{\rm 5}$BOLD}
\footnote{CO5BOLD stands for ``COnservative COde for the COmputation of 
        COmpressible COnvection in a BOx of L Dimensions with L=2, 3.'' 
        In this work we use L=3.}
code
\citep{2012JCoPh.231..919F}.
The code solves the time-dependent non-linear magnetohydrodynamics (MHD)
equations in a Cartesian box with an external gravity field, taking
non-grey radiative transfer into account.

We study four models representing different regions of the solar surface
convection. In each of the models, the computational domain is 
{$38.4\times38.4\times2.8$~Mm$^3$} in the $x \times y \times z$
direction, discretised on {$480\times480\times120$} grid cells.
The cell size in the horizontal is 80~km and the vertical cell size
varies from 50~km in the lower part of the computational domain
(convection zone) down to 20~km near the surface and in the atmosphere.
The domain extends $\sim$1.3~Mm above the mean Rosseland optical depth
$\tau_{R}$=1, providing the atmosphere wherein the waves propagate. The
non-magnetic run (Sun-v0) is computed by setting the initial magnetic
flux density to zero. The three magnetic runs, namely Sun-v10, Sun-v50,
and Sun-v100, are computed by embedding the initial model with a uniform
vertical field of 10, 50, and 100~G, respectively, in the entire domain.
A detailed summary of the numerical models used in this work is described in
\citetalias{2019ApJ...872..166V}. 

Periodic boundary conditions are used for the side boundaries in all
four models. The velocity field, radiation, and the magnetic field
components are periodic in the lateral directions. The top boundary is
open for fluid flow and outward radiation, with the density decreasing
exponentially in the boundary cells outside the computational domain. The vertical
component of the magnetic field is constant across the upper boundary
and the transverse component drops to zero at the boundary. The bottom
boundary is setup in such a way that the in-flowing material carries a
constant specific entropy resulting in a radiative flux corresponding to
an effective temperature ($T_\mathrm{eff}$) of $\sim$5770~K. The bottom
boundary conditions for the magnetic fields are the same as for the top
boundary.

An equation of state that adequately describes the solar plasma,
including partial ionisation effects, is used. The
radiative transfer proceeds via an opacity binning method adapted from the MARCS stellar atmosphere
package
\citep{2008A&A...486..951G}, with the help
of five opacity groups. 
For the radiative transfer we use long characteristics along a set of
eight rays.

\section{Synthetic spectra}\label{s:synthetic_spectra}
We use the publicly available open-source, radiative transfer code
NICOLE\footnote{https://github.com/hsocasnavarro/NICOLE}
\citep{2015A&A...577A...7S}
to compute synthetic spectra in local thermodynamic equilibrium (LTE).
The code solves the equation of radiative transfer along one-dimensional 
(1D) vertical columns assuming a plane-parallel atmosphere. NICOLE is 
capable of non-LTE (NLTE) radiative transfer calculations, but we are 
mainly interested in the properties of the waves propagating in the low 
photosphere where the assumption of LTE is still valid. However, a 
proper treatment of the upper layers will require NLTE analysis, which 
we defer to a future study.

The 3D snapshots from the numerical simulation are interpolated to a
constant $z$-grid with $\Delta z$=10 km. The density, gas pressure, and
velocity on the corresponding Rosseland mean optical depth grid are then
passed as the model atmosphere to the synthesis code. The code uses its
native equation of state (EOS) to calculate the electron pressure. We use the same set of abundances that are used in the 
{CO$^{\rm 5}$BOLD} simulation. The radiative transfer is then carried 
out for each 1D column (480$\times$480) in the horizontal direction 
independently and for each snapshot. The spectral lines are sampled at 
5~m\AA{} resolution spanning a wavelength range of $\pm500$~m\AA{} 
around the central wavelength.

\subsection{Line selection}\label{ss:line_selection}
\begin{figure}[ht!]
        \centering
        \includegraphics[width=1.\columnwidth]{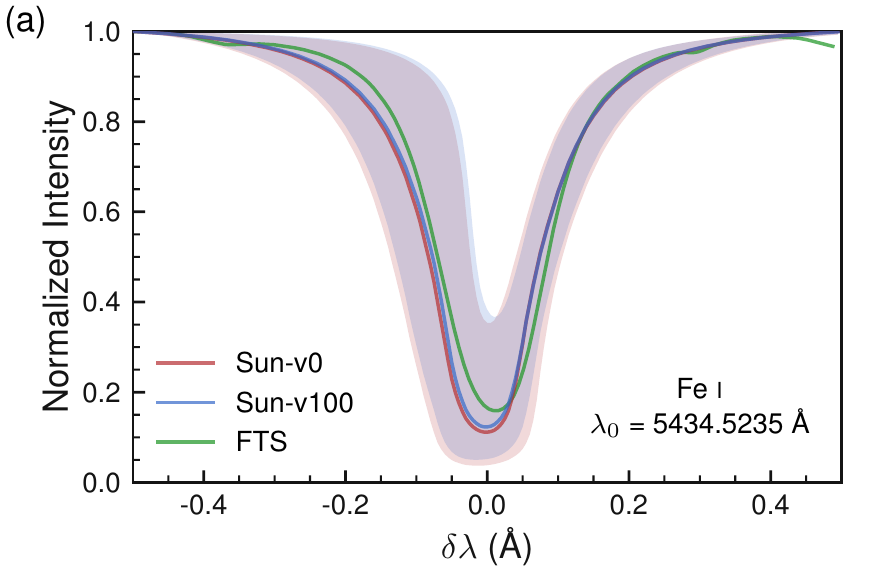}\\
        \includegraphics[width=1.\columnwidth]{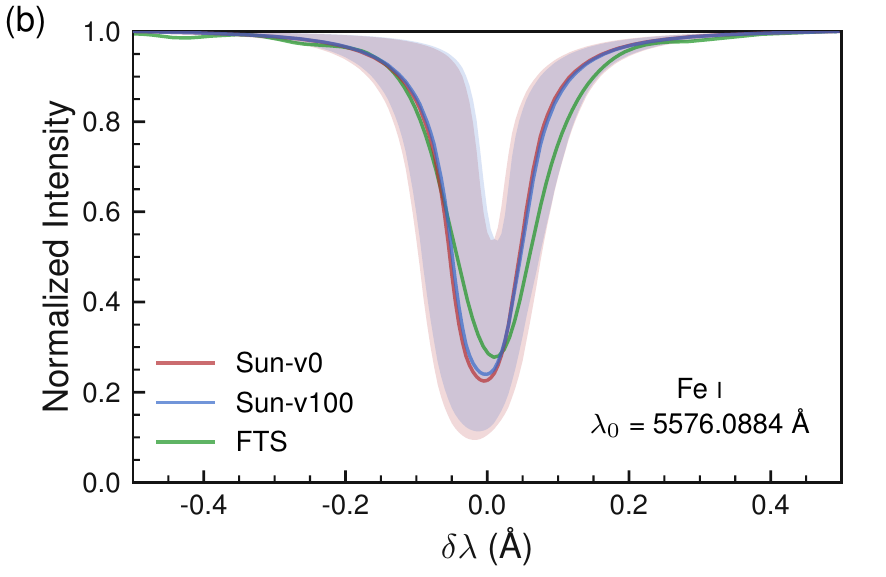}
        \caption{Median profile of (a) \ion{Fe}{I}~$\lambda$~5434~\AA~and 
                (b) \ion{Fe}{I}~$\lambda$~5576~\AA~taken from a snapshot of the 
                Sun-v0 (blue) and Sun-v100 (red) models. The FTS atlas spectra 
                for the respective lines are shown in green. The background 
                shaded region shows the 95\% bounds of the line profiles over the 
                whole computational domain in the selected snapshot.}
        \label{fig:profiles}
\end{figure}

We consider two magnetically insensitive neutral iron lines 
{\ion{Fe}{I}}~$\lambda\lambda$~5434~\AA{} and 5576~\AA{}, both of which have a 
Land\'{e} $g$-factor equal to zero. The atomic parameters of our selected lines 
are listed in Table~\ref{tab:line_params}. Figure~\ref{fig:profiles}
shows the average profile of the selected lines from a snapshot of the
non-magnetic (Sun-v0) and one of the magnetic models (Sun-v100) along with the 
Fourier transform spectrometer (FTS) atlas spectra for the lines. The 
95\% bounds of the spectral profile over the whole domain in the 
selected snapshot are also shown. Our choice of the spectral lines is 
based on the work by
\cite{2011A&A...532A.111K}, 
where these two \ion{Fe}{I} lines were used to estimate the energy
transport by acoustic and atmospheric gravity waves in the quiet Sun.
\begin{table}[ht!]
        \caption{Atomic parameters used for the synthesis of the spectral maps}             
        \label{tab:line_params}      
        \centering          
        \begin{tabular}{l c c }     
                & \ion{Fe}{I}~$\lambda$~5434~\AA{} & \ion{Fe}{I}~$\lambda$~5576~\AA{}\\
                \hline            
                Central wavelength (\AA{})& 5434.5235 & 5576.0884 \\
                Excitation potential (eV) & 1.0110557 & 3.43019 \\
                log($gf$) & -2.122 & -0.94 \\
                Transition & $^5$F$_{1}$ $-$ $^5$D$_{0}$ & $^5$F$_{1}$ $-$ $^5$D$_{0}$ \\   
                Collisions &  Barklem &  Barklem \\
                Damping sigma & 243 & 854 \\
                Damping alpha & 0.247 & 0.232 \\
                \hline                  
        \end{tabular}
\tablefoot{The spectroscopic data of the selected lines are taken from 
the National Institute of Standards and Technology (NIST) Atomic Spectra Database 
\citep{NIST_ASD} 
and the data for collisional broadening by hydrogen are from 
\cite{1998PASA...15..336B}.
}
\end{table}

The two lines differ mainly in their temperature sensitivity; the
low-excitation potential {\ion{Fe}{I}~$\lambda$~5434~\AA{}} line is more
temperature sensitive than the {\ion{Fe}{I}~$\lambda$~5576~\AA\ line{}}, the latter being otherwise considered to be temperature insensitive
\citep{2005A&A...439..687C}. 
The line-forming surface of the {\ion{Fe}{I}~$\lambda$~5434~\AA{}} line is
therefore less smooth than that of the {\ion{Fe}{I}~$\lambda$~5576~\AA{}}
line; this has a more adverse effect on the phase difference analysis
that is used to investigate waves, a point to which we come later in
Sect.~\ref{s:results}.

\subsection{Synthetic Dopplergrams and velocity response function}
\label{ss:method_dopplergram}
To study IGWs, we primarily examine the 2D velocity field in the solar
atmosphere, which can be readily obtained from the Doppler shift of
observed spectral lines.  The major drawback is that they provide an
average of the velocity field over a broad height range, effectively
producing a smeared out form of the velocity perturbations caused by any
passing wave. Nevertheless, the line shifts can be estimated using
different techniques from the same line profile which provide velocity
measurements corresponding to different atmospheric layers. In this
work, we estimate three different Doppler velocities from the
synthesised  spectral lines. Our first approach is to use the Fourier
method, where the phase of the first Fourier component of the line
profile gives us the Doppler shift. In the second approach, we estimate
the velocity from the shift in the centre of gravity (COG) of the line.
Since these two approaches look at the overall shift in the spectral
line, they give similar values in terms of where the velocity signal is
formed. In this paper, we mainly discuss results from the COG
method. Lastly, we compute the shift of the line profile minima by
fitting a second-degree polynomial in the range $\pm$30m\AA{} around the
minimum position of the spectral line. The line minima method gives us
the velocity field in the higher layers when compared to the Fourier or
the COG method. Some of the aforementioned methods have been discussed
in the literature
\citep{1995SoPh..162..129S,
          2011SoPh..271...27F,
          2012SoPh..278..217C,
          2014SoPh..289.3457N}.

\begin{figure}[ht!]
        \centering
        \includegraphics[width=1\columnwidth]{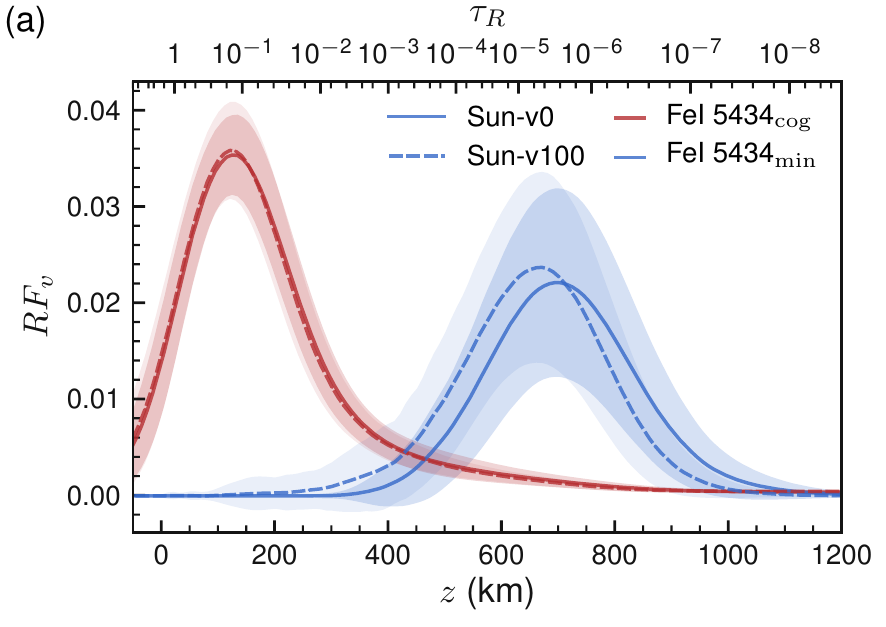}
        \includegraphics[width=1\columnwidth]{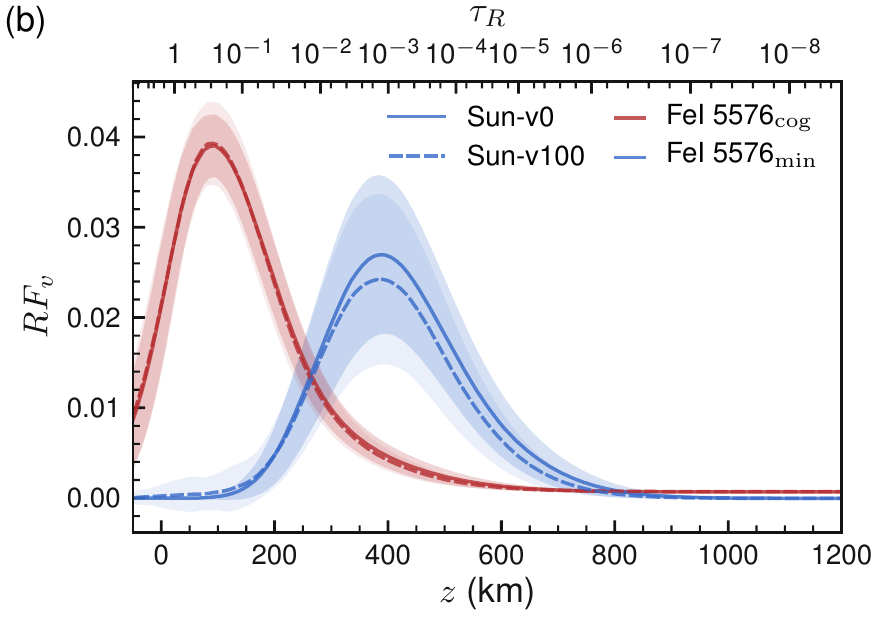}
        \caption{Mean $RF_{v}$  of the velocities 
                determined using the COG shift (red) and the line-core shift 
                (blue) for the
                (a) {\ion{Fe}{I}~$\lambda$~5434~\AA{}} and 
                (b) {\ion{Fe}{I}~$\lambda$~5576~\AA{}} lines in the non-magnetic model (Sun-v0; solid) and in 
                a magnetic model (Sun-v100; dashed). The spread of the 
                $RF_{v}$ ranging from an inter-granular to granular  region is 
                represented by the background shade with the fainter shade 
                representing the magnetic model.}
        \label{fig:response_profiles}
\end{figure}

\begin{figure*}[ht!]
        \centering
        \includegraphics[width=1.\textwidth]{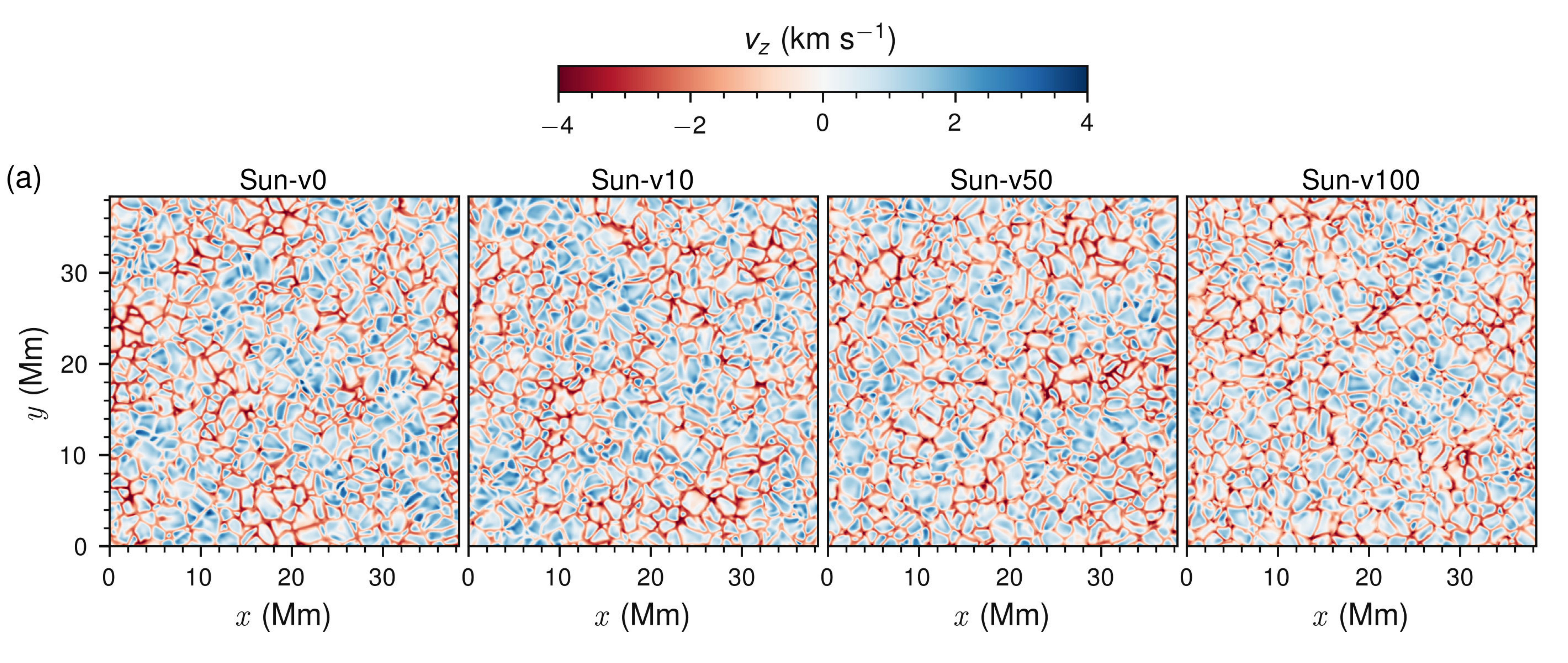}
        \includegraphics[width=1.\textwidth]{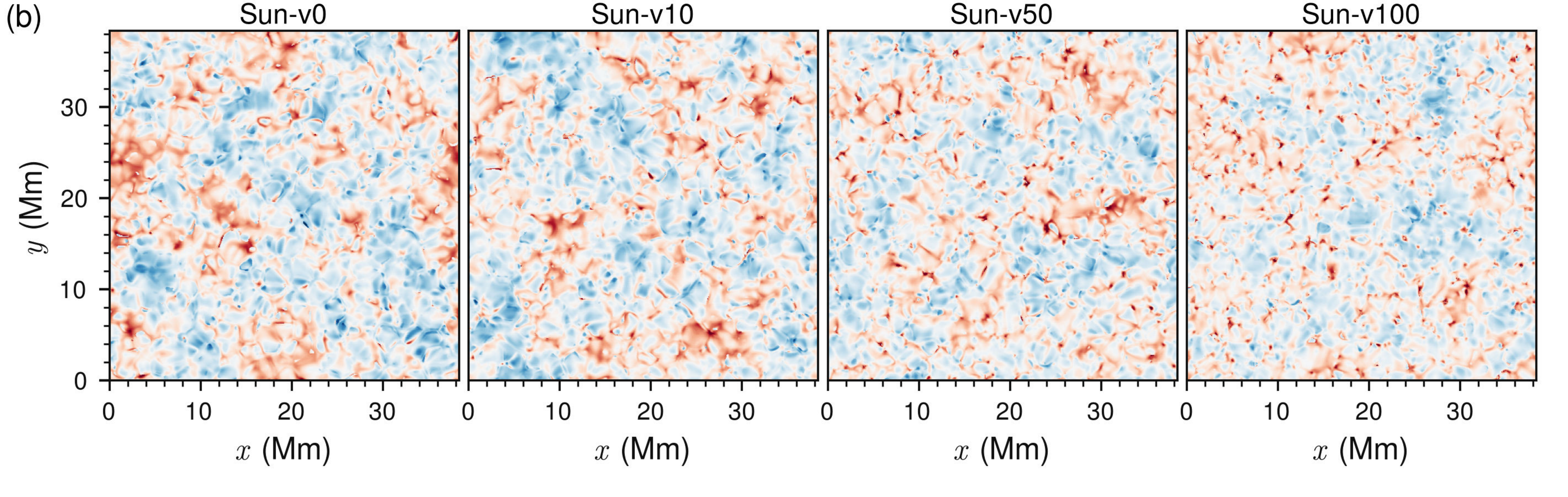}
        \caption{Snapshot of Dopplergrams from 
                \ion{Fe}{I}~$\lambda$~5576~\AA{} for the four different models 
                (left to right: increasing average magnetic flux density from 
                0 to 100~G), estimated using (a) the COG method 
                which provides the velocity field around $z=$100~km and (b) the 
                line-minima shift which provides the velocity field around 
                $z=$400~km}
        \label{fig:dopplergrams}
\end{figure*}

We now look at the velocity response function ($RF_{v}$) to estimate the
height of formation of the velocity signals obtained from the different
methods described above. The response functions are calculated using the
method described in
\cite{2001A&A...371.1128E}. 
In Fig.~\ref{fig:response_profiles}(a), we compare the mean $RF_{v}$
for the {\ion{Fe}{I}~$\lambda$~5434~\AA{}} line in the non-magnetic model (Sun-v0)
and in one of the magnetic models (Sun-v100), both obtained using the
COG method and the line minima method. The spread of the $RF_{v}$
ranging from the inter-granular to the granular region is represented by the
shaded background with the fainter shade representing the magnetic
model. The corresponding profiles for the
{\ion{Fe}{I}~$\lambda$~5576~\AA{}} are shown in
Fig.~\ref{fig:response_profiles}(b). We see that the velocity obtained
using the COG method is formed mainly close to the surface around
$z=$100. There is a small difference in the height of formation for the
two lines of $\sim$30~km,  with the
\ion{Fe}{I}~$\lambda$~5434~\AA{} line forming slightly higher up in the
atmosphere. We do not see any difference in formation height between the
non-magnetic and the magnetic models. The line-minima method on the
other hand probes the higher layers, as evident from the corresponding
$RF_{v}$ (blue curves) shown in Fig.~\ref{fig:response_profiles}. The peak of the
$RF_{v}$ for the two \ion{Fe}{I} lines is separated by 280--310~km,
with the magnetic models peaking lower than their non-magnetic
counterpart in the case of {\ion{Fe}{I}~$\lambda$~5434~\AA{}}. The
$RF_{v}$ for the {\ion{Fe}{I}~$\lambda$~5576~\AA{}} line using the COG
method that we computed using our simulation models matches that
of
\cite{2002A&A...381..253B}. 
The $RF_{v}$ for the line-minima in our case is formed 50~km above
compared to
\cite{2002A&A...381..253B} 
and approximately at the same height as in the case of 
\cite{2009A&A...508..941B}. 
We also compared the $RF_{v}$ for the \ion{Fe}{I}~$\lambda$~5434~\AA{}
line with that of
\cite{2010A&A...522A..31B} 
and 
\cite{2011A&A...532A.111K} 
and see that the line-forming peaks are systematically shifted higher up
in our case by around 100~km. The discrepancies may be due to the
absence of a chromospheric temperature rise in our models or due
to the assumption of LTE. A representative snapshot of the Dopplergrams for
the \ion{Fe}{I}~$\lambda$~5576~\AA{} line obtained using the COG method
and the line-minima method is shown in Fig.~\ref{fig:dopplergrams}
showing the difference in the velocity field as a result of probing
different layers using the same spectral line.

\section{Analysis}\label{s:analysis}

Now that we have computed the different Doppler velocities and estimated the
height of formation of the velocity signals for each line, we can now
look for signatures of waves in the velocity field. We carry out a
spectral analysis of the 2D synthetic observations by
Fourier-transforming the physical quantities in both space and time to
identify and separate IGWs from other types of waves present in the
domain. The analysis methodology is already presented in
\citetalias{2017ApJ...835..148V} 
and 
\citetalias{2019ApJ...872..166V}. 
In the foregoing work, we investigated the wave signatures from the
velocity signals obtained directly from the simulation box. The analyses
was carried out at constant geometrical height, thereby allowing us to
study the wave generation and propagation as it happens in the physical
volume. In this work, we are nevertheless interested in the effect of probing
layers of constant optical depth scale (through information provided by
spectral lines) on the inferred properties of the waves. The levels of
constant optical depth are highly corrugated in the atmosphere
and therefore it is still unclear what effect they may have on the
observations of waves. Therefore, it is important to study the
behaviour of velocities at constant (Rosseland) optical depth
interpolated directly from the simulations and to see if one is able to
still recover the signature of IGW propagation and the effect of
magnetic field. This will enable us to properly interpret the
spectral signatures obtained from the estimates of velocities derived
from the spectral lines. We investigate the velocity--velocity ($v-v$)
phase difference and the related coherence spectra obtained from
different velocity measurements with the aim of comparing them with
observations.

In this work, we use three-hour time-series of the synthetic maps and
the domain spans 38.4~Mm in the horizontal direction. This gives us a
frequency resolution ($\Delta\nu$) of 92.6~$\mu$Hz and a wavenumber
resolution ($\Delta k_{h}$) of 0.164 rad Mm$^{-1}$. The grid resolution
of 80~km results in a Nyquist wavenumber ($k_{Ny} = \pi/\delta x$) of
39.25~rad~Mm$^{-1}$ of which we are only interested in horizontal
wavenumbers up to 10~rad~Mm$^{-1}$, where the bulk of IGWs occur. A
vertical and horizontal grid constant of, respectively 20 and 80~km is
sufficient to capture the range of the internal wave spectrum in the
models as discussed in
\citetalias{2017ApJ...835..148V}. 
Snapshots from the simulations are taken at 30 s intervals, and the
spectral line synthesis is performed for each snapshot, resulting in a
Nyquist frequency ($\nu_{Ny}$) of 16.66~mHz.

\subsection{Zero coherence threshold}\label{ss:zero_coherence_threshold}
\begin{figure}[h!]
        \centering
        \includegraphics[width=1.\columnwidth]{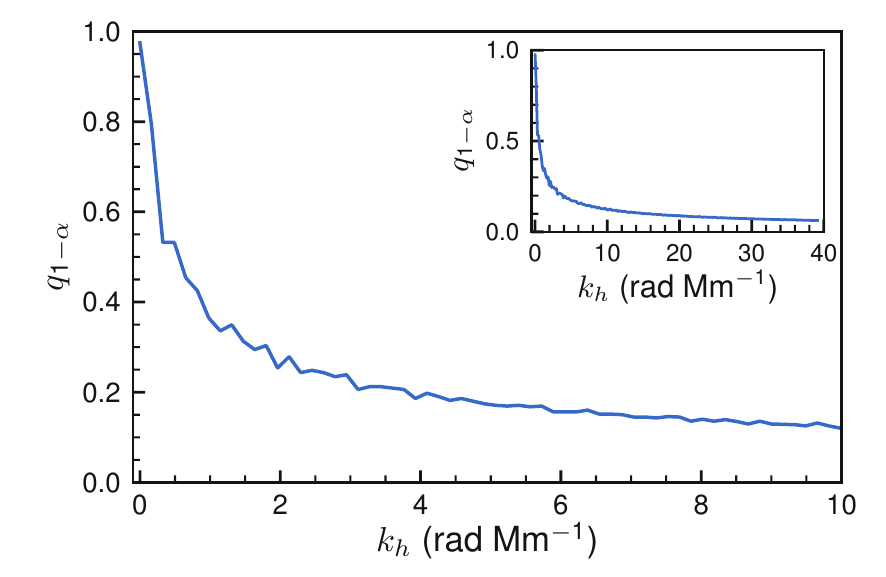}\\
        \caption{Zero-coherence threshold as a function of $k_{h}$ at the 
                significance level $\alpha$=0.05. The inset plot shows the 
                entire range up to the Nyquist wavenumber 
                ($k_{Ny}$ =39.25~Mm$^{-1}$).}
        \label{fig:cohthreshold}
\end{figure}

Coherence measures the linear dependence of the two processes, which in
our case is the velocity signals at two different heights. However,
even for two independent processes, the coherence estimator will give a
non-zero value due to fact that the spectra are computed by averaging of
several individual segments. We consider a confidence threshold with a
certain probability such that if the estimated coherence is below this
threshold, zero coherence is indicated for that Fourier component.
Following
\cite{HannanBook}, 
the zero coherence threshold ($q_{1-\alpha}$) at the significance level
$\alpha$ can be expressed as
\begin{equation}
q_{1-\alpha} = \sqrt{1-\alpha^{2/(N_{s}-2)}},
\end{equation} 
where $N_{s}$ is the equivalent number of degrees of freedom in the case
of independent segments. If the estimated coherence $\mathcal{K} >
q_{1-\alpha}$, the null hypothesis of zero coherency can be rejected
with a 100$\alpha$\% probability of error. In our case, $N_{s}$  can be
taken as the number of averages performed for a given $k_{h}$ in the
$k_{x}-k_{y}$ plane. Figure~\ref{fig:cohthreshold} shows the zero
coherence threshold as a function of $k_{h}$ at the significance level
$\alpha$=0.05. The threshold decreases with $k_{h}$ due to the
increasing number of averages in the $k_{x}-k_{y}$ plane and falls below
0.5 for $k_{h}>$1~rad~Mm$^{-1}$. The uneven curve is the result of
rasterisation in the $k_{x}-k_{y}$ plane.

\subsection{Confidence interval for the phase and coherence}
\label{ss:conf_interval}
In this study, we also consider confidence intervals for the measured
phase and coherence. The 100(1-$\alpha$)\% confidence interval for the
estimated phase ($\hat{\phi}$) is given by the inequality
\citep{HannanBook}
\begin{equation}
    |\sin(\hat{\phi} - \phi)| \leq \left\{ \frac{1-\mathcal{K}^2}{\mathcal{K}^2 (2N_{s}-2)} \right\}^{1/2} t_{2N_{s}-2} \left( \frac{\alpha}{2} \right),
\end{equation}
where $t_{2N_{s}-2} \left( \frac{\alpha}{2} \right)$ is the
$(1-\alpha/{2})^{\rm th}$ quantile of the $t$-distribution with
$2N_{s}-2$ degrees of freedom.

An approximation for the 100(1-$\alpha$)\% confidence interval for the
estimated coherence ($\mathcal{K}$) is given as
\citep{HannanBook}
\begin{equation}
\displaystyle q_{l,u} = \tanh\left\{\tanh^{-1}(\mathcal{K}) \pm (u_{\alpha/2}) (2(N_s - 1))^{-1/2} - (2(N_{s}-1))^{-1} \right\},
\end{equation} 
where $u_{\alpha/2}$ is the $(1-\alpha/{2})^{\rm th}$ quantile of the
standard normal distribution.

\subsection{Energy flux estimation}\label{ss:analysis:energy_flux}
The energy flux of the IGWs can be directly determined from the
simulation as described in 
\citetalias{2017ApJ...835..148V}
and 
\citetalias{2019ApJ...872..166V}. 
However, in this study we are interested in the diagnostic information
provided by the observed spectral lines and in determining the energy
flux using the phase difference of the Doppler velocity measurements.

Given the difference between the formation heights ($\Delta z$) of the
two lines, the vertical phase velocity spectrum is determined from the
phase difference spectra as
\begin{equation}
        v_{{\rm ph},z} = \Delta z /\Delta t,
\end{equation}
where $\Delta t = \Delta \Phi/\omega$ represents the vertical phase
travel time between the two layers provided that the medium through
which the wave propagates is homogeneous. Here, $\Delta\Phi$ represents
the phase difference in radians between the two layers for a harmonic
wave of frequency $\omega$. The vertical group velocity spectrum is then
determined from the phase velocity as
\citep{2008ApJ...681L.125S,
           2011A&A...532A.111K}
\begin{equation}
v_{{\rm g},z} = - v_{{\rm ph},z} \sin^2{\theta},
\end{equation}
where $\theta$ is the angle between the horizontal and the wavevector
${\boldsymbol{k}}$, given using the dispersion relation as
$\cos{\theta}=\omega/N$ in the incompressible limit, with $N$
representing the Brunt-V\"{a}is\"{a}l\"{a} frequency. Finally, we obtain
the vertical energy flux spectrum as the product of energy density and
vertical group velocity,
\begin{equation}
\mathcal{F}_{z} = \frac{1}{2} \varrho \langle v^2 \rangle v_{{\rm g},z}.
\end{equation}

\section{Results and discussion}\label{s:results}
We now present results from our spectral analysis. First, we analyse the
intensity maps and compute the power, phase, and coherence on the
$k_{h}-\omega$ diagnostic diagram. We then turn to the analysis of the
velocity data, presenting first the results from constant optical depth
and then those of the spectral analysis of the synthetic
Dopplergrams. We compare the energy flux estimation from the constant
optical depth measurements and the Dopplergrams.

\subsection{Power, phase difference, and coherence spectra of intensity}
\begin{figure*}[h!]
        \centering
        \includegraphics[width=1.\textwidth]{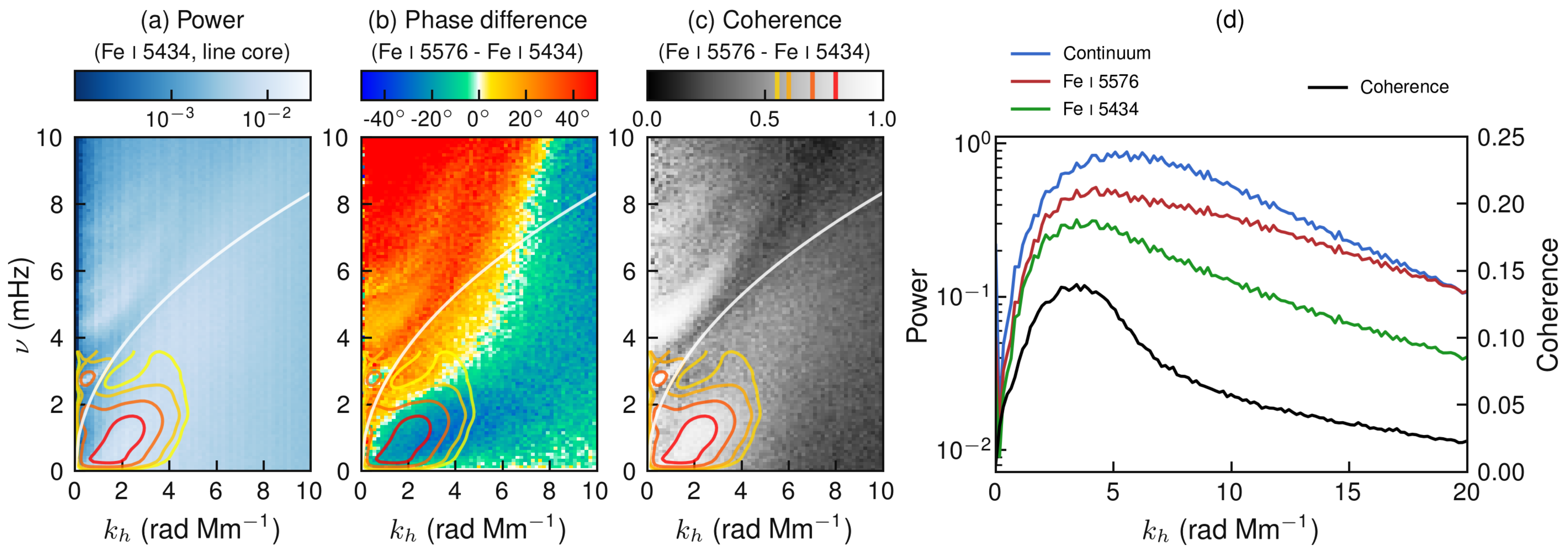}
        \caption{
                        (a) Power in the $k_{h}-\omega$ diagnostic diagram of the 
                        line core intensity of the {\ion{Fe}{I}~$\lambda$~5434~\AA{}} 
                        line. The phase (b) and coherence (c) spectra between the 
                        line-core intensity of the {\ion{Fe}{I}~$\lambda$~5576~\AA{}} 
                        and {\ion{Fe}{I}~$\lambda$~5434~\AA{}} lines are also shown. 
                        The white line marks the dispersion relation of 
                        the surface gravity waves (\textit{f}-mode). 
                        The coloured contour lines in (a)-(c) mark the coherence 
                        levels with the reference colours shown in the colourmap of 
                        (c). (d) Power as a function of $k_{h}$ integrated over 
                        the IGW region for the continuum intensity and line-core 
                        intensities of {\ion{the Fe}{I}~$\lambda\lambda$~5576~\AA{}} 
                        and {5434~\AA{}} lines. Also shown in black is the mean 
                        coherence in the IGW region.}
        \label{fig:intensity_intensity_ph_ch}
\end{figure*}

As an illustration of the importance of multi-spectral observations for
the study of IGWs, we first look at the diagnostic spectra of intensity
obtained from the synthetic observations of the non-magnetic model
(Sun-v0). In Fig.~\ref{fig:intensity_intensity_ph_ch}, the leftmost
plot shows the $k_{h}-\omega$ power spectra of the line core intensity
of the {\ion{Fe}{I}~$\lambda$~5434~\AA{}} line. The second and third plots
show the phase and coherence spectra, respectively, between the line
core intensities of {\ion{Fe}{I}~$\lambda$~5576~\AA{}} and
{\ion{Fe}{I}~$\lambda$~5434~\AA{}}. The rightmost plot shows the power
as a function of $k_{h}$ integrated over the IGW region for the
continuum intensity and line core intensities of
{\ion{Fe}{I}~$\lambda\lambda$~5576~\AA{}} and {5434~\AA{}} lines. The
mean coherence over the IGW region as a function of $k_{h}$ is shown in
black. It is clear from the power spectra that, whether it be the
continuum intensity which is formed low in the atmosphere, or the line
core which is formed in the higher layers, the power in the IGW region
is dominated by convective velocity signal from the super granular
scales ($k_{h}\lessapprox1$~rad~Mm$^{-1}$) to the granular scales
($k_{h}\gtrapprox2$ rad Mm$^{-1}$) . The signature of the coherent IGW
waves ($k_{h} \approx 1 -  4$~rad~Mm$^{-1}$) with their characteristic
downward phases (see, for e.g.
\citetalias{2017ApJ...835..148V}) 
cannot be discerned in the power spectra. This is evident from the
location of the enhanced coherence in
Fig.~\ref{fig:intensity_intensity_ph_ch}(d) relative to the power
spectra. To make this point clear, we over-plotted contour lines of
four coherency values greater than 0.5 on the power spectra and the
phase spectra in Figs.~\ref{fig:intensity_intensity_ph_ch}(a)-(c). An
internal wave transporting energy at an angle to the vertical, with an
upward component, will have a downward-propagating phase component that
shows up as a negative phase lag between two geometrical heights. The
negative phases (blue-green) within the contour clearly show that this
region marks the presence of upward propagating IGWs, which are
otherwise indistinguishable in the power spectra. This highlights the
importance of simultaneous or quasi-simultaneous 2D narrowband imaging
by scanning over a single line or multiple lines in order to study IGWs
in the solar atmosphere. Our focus in the subsequent sections is on
the analysis of the velocity measurement obtained from the two
{\ion{Fe}{I}} lines.

\subsection{Phase and coherence spectra at constant optical depth}
\begin{figure*}[ht!]
        \centering
        \includegraphics[width=.98\columnwidth]{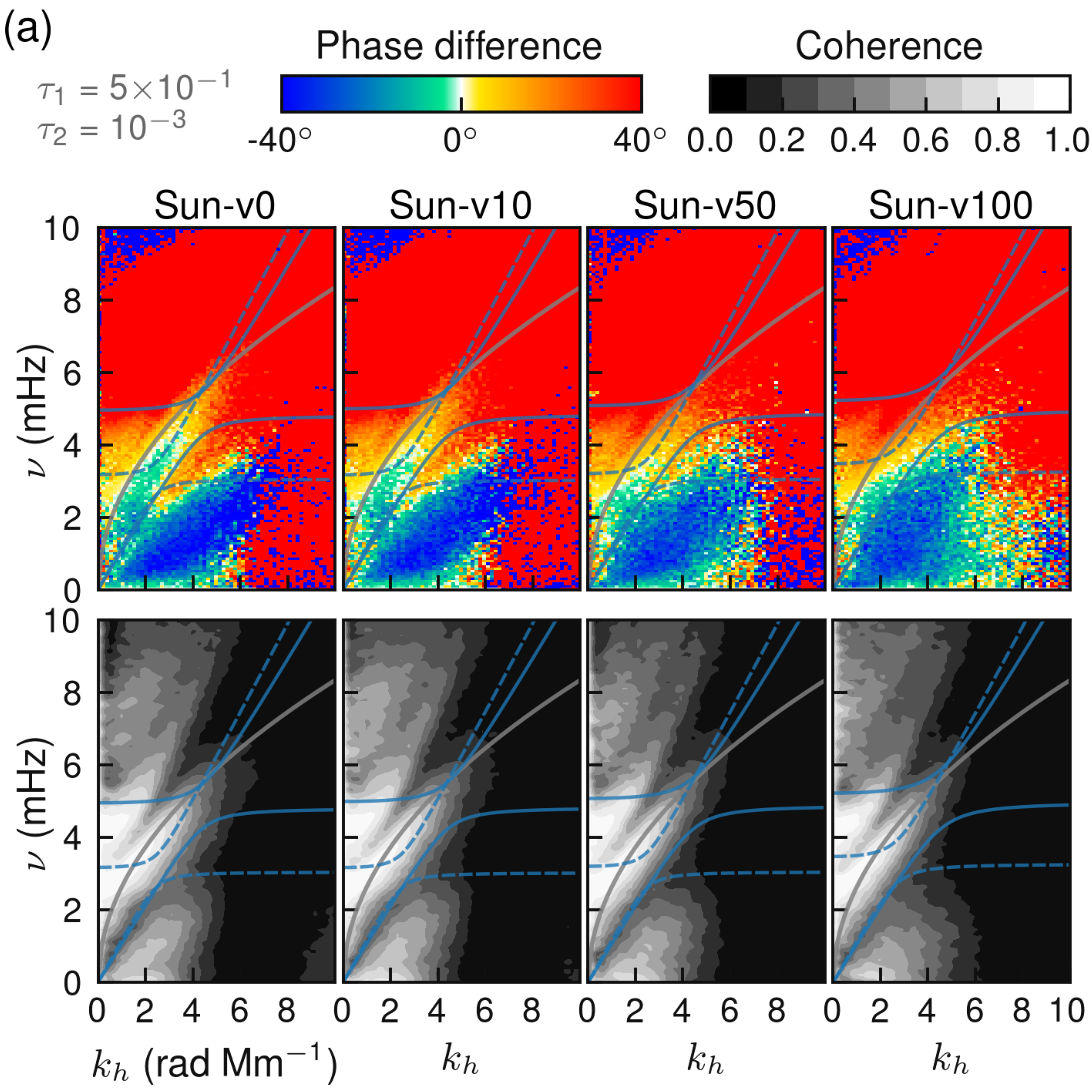}\hskip3ex\includegraphics[width=.98\columnwidth]{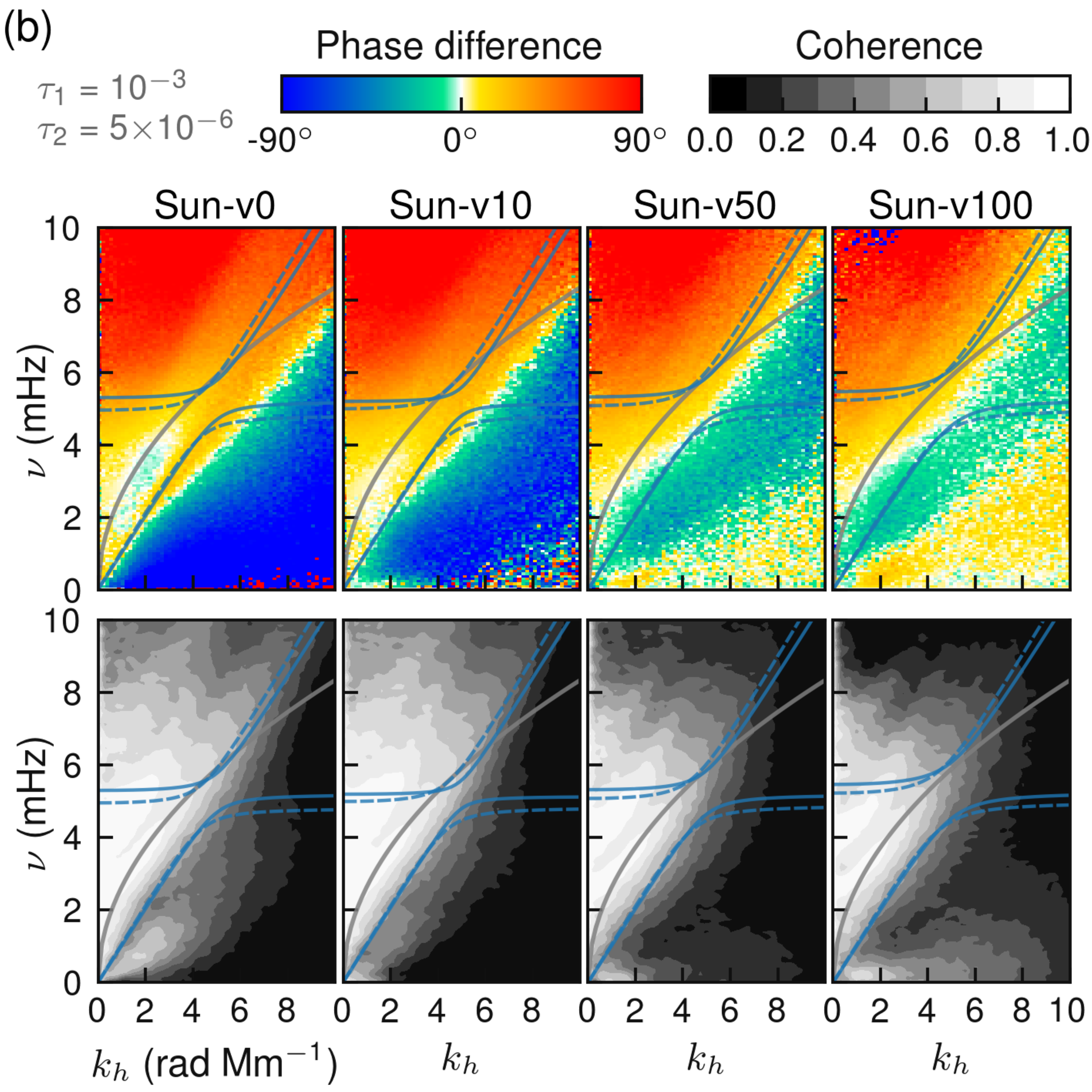}
        \caption{Phase difference (colour) and coherence (greyscale) spectra 
                between layers of constant optical depth (a) $\tau_{R}=0.5$ 
                and $1.0\times10^{-3}$ and (b) $\tau_{R}=1.0\times10^{-3}$ and 
                $5.0\times10^{-6}$. The grey curve marks the dispersion 
                relation of the surface gravity waves. The blue curves mark the 
                propagation boundary of acoustic-gravity waves for the 
                top (solid) and bottom (dashed) layers.}
        \label{fig:ph_ch_spectra_tau}
\end{figure*}

We start with the velocity measurements interpolated from the simulation
domain on to surfaces of constant mean (Rosseland) optical depth. In
Fig.~\ref{fig:ph_ch_spectra_tau}, we show the phase and coherence
spectra between two pairs of velocity measurements each taken at a
constant optical depth layer for all the four models studied here. We
select three representative layers, namely $\tau_{R}=0.5,
1.0\times10^{-3}, 5.0\times10^{-6}$. These correspond to the layers
close to where the different velocity response  functions for the two
lines show a peak as shown in Fig.~\ref{fig:response_profiles} in
Sect.~\ref{s:synthetic_spectra}.

We first look at the phase and coherence spectra for the pair
$\tau_{R}=0.5$ and $1.0\times10^{-3}$ in 
Fig.~\ref{fig:ph_ch_spectra_tau}(a). The blue curves mark the
propagation boundary of acoustic-gravity waves for the top (solid) and
bottom (dashed) layers, with the IGWs located below the lower
propagation boundary. Here we use the isothermal cut-off
frequencies
\citepalias[see][]{2017ApJ...835..148V}.
Although all the four models show a downward phase propagation in the
IGW regime, an apparent difference in the phases due to the effect of
magnetic field is evident when we compare the four different models. The
decrease in the phase difference in the IGW regime is probably due to
the decreasing geometrical height difference between the two constant
$\tau_{R}$ layers with the increasing average magnetic field. We also
notice that beyond $k_{h}=8$~rad Mm$^{-1}$ the phase spectra become more
noisy in all the models, which provides hints about the spatial scales over which the
optical depth surface may show drastic fluctuations. However, we see
from the coherence spectra for this pair of layers that the coherence
is close to 0 for $k_{h}>5$~rad Mm$^{-1}$. Therefore, we may deduce that
a reliable estimate of the phase spectra can only be taken for $k_{h}$ 
up to 5~rad Mm$^{-1}$. Tracing the dispersion relation curve of the surface gravity waves ($f$-mode)  shown in grey reveals an apparent phase
difference in the non-magnetic model that is completely absent from
stronger magnetic models.

Looking now at the phase and coherence spectra for the pair
$\tau_{R}=1.0\times10^{-3}$ and $5.0\times10^{-6}$ shown in
Fig.~\ref{fig:ph_ch_spectra_tau}(b), we indeed see the signature of upward
phase propagation in the strong magnetic model as a result of downward-propagating waves. However, this downward propagation is rather
restricted to the region below $\nu=2$~mHz. In the region above $\nu=2$~mHz (within
the IGW regime) we notice that the phases are positive, but the
coherence spectrum at these locations is below the zero coherence
threshold, and therefore the phase measurements are unreliable. The surface
gravity waves ($f$-mode) show an almost zero phase difference in the
non-magnetic model but are, like in the lower layers, completely absent
in stronger magnetic models.

\subsection{Phase and coherence spectra from Doppler velocities}
Now we turn to the analysis of wave spectra from the Doppler velocities
estimated from the synthetic spectral lines. First, we study the phase
and coherence spectra between Doppler velocities from a single line
employing the different definitions of the Doppler velocity described in
Sect.~\ref{ss:method_dopplergram}. We compare the line-minima and
COG method which gives us the velocity field at different layers in the
atmosphere. This is followed up by looking at the phase and coherence
spectra of two separate lines utilising either the COG shift or the 
line-core Doppler shifts.

\begin{figure}[h!]
        \centering
        \includegraphics[width=.98\columnwidth]{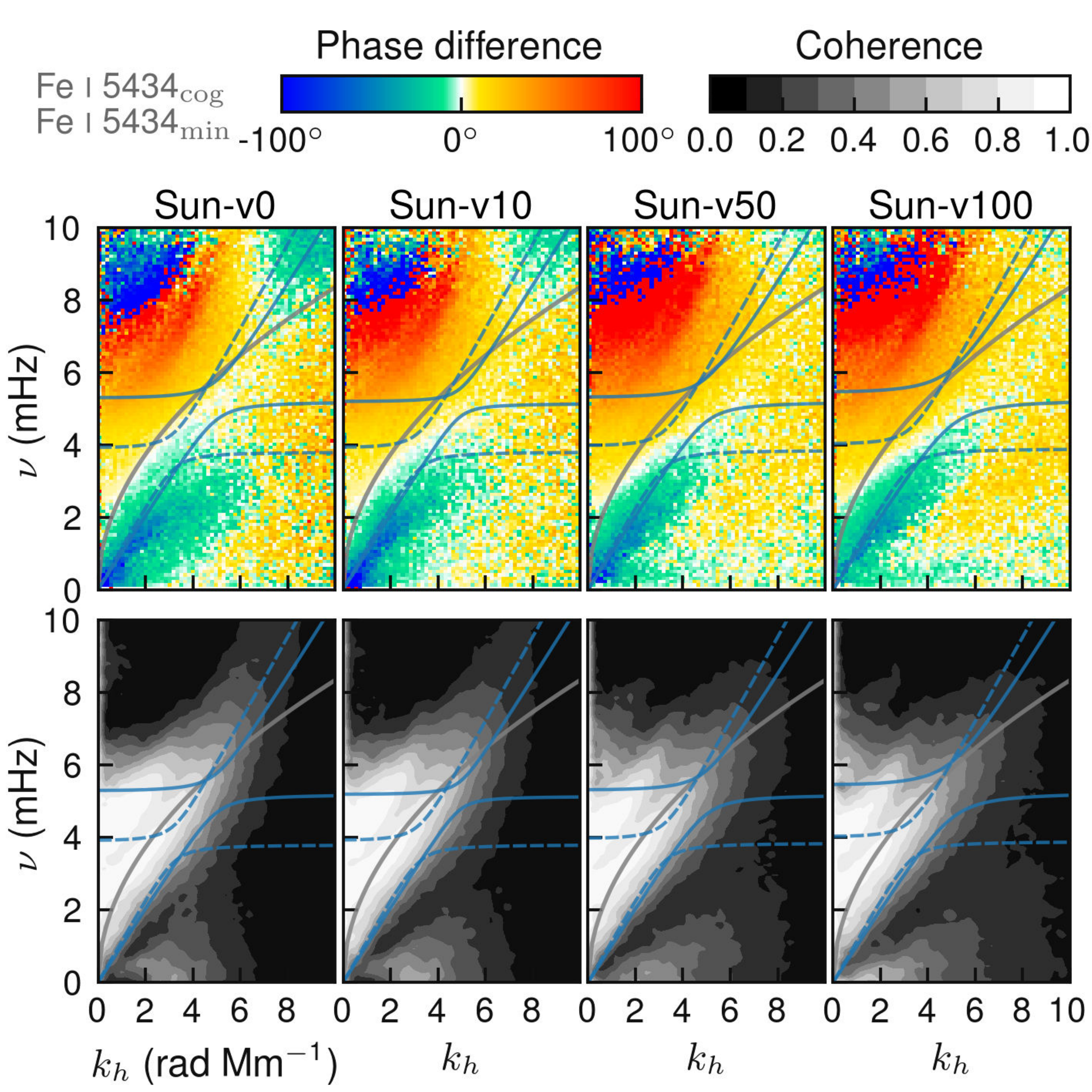}
        \caption{Phase difference (colour) and coherence (greyscale) spectra 
                for {\ion{Fe}{I}~$\lambda$~5434~\AA{}} line measured with the 
                velocities derived using the COG and line-minima shifts. The 
                grey curve marks the dispersion relation of the surface gravity 
                waves. The blue curves mark the propagation boundary of the 
                acoustic-gravity wave corresponding to the approximate formation 
                heights where the COG (dashed) and line-minima (solid) velocities 
                are measured.}
        \label{fig:ph_ch_spectra_single_5434}
\end{figure}

In Fig.~\ref{fig:ph_ch_spectra_single_5434}, we show the phase
and coherence spectra for the {\ion{Fe}{I}~$\lambda$~5434~\AA{}} line.
Here, the Doppler velocities obtained using the COG method and the
line-minima method are used for computing the phase and coherence
spectra. According to the $RF_{v}$ shown in
Fig.~\ref{fig:response_profiles}(a) the two velocity signals are
separated by around {550--570}~km. As a result of this large separation
the coherency is very low, especially in the IGW region. The effect of
magnetic field is negligible in the phase spectra, with the Sun-v100
model showing a marginal upward phase propagation (see area covered in
yellow within the IGW regime compared to the Sun-v0 model) for
wave numbers larger than $k_{h}$ = 4~rad~Mm~$^{-1}$. However, the low
coherence in this region renders any phase estimate unreliable.

\begin{figure}[h!]
        \centering
        \includegraphics[width=.98\columnwidth]{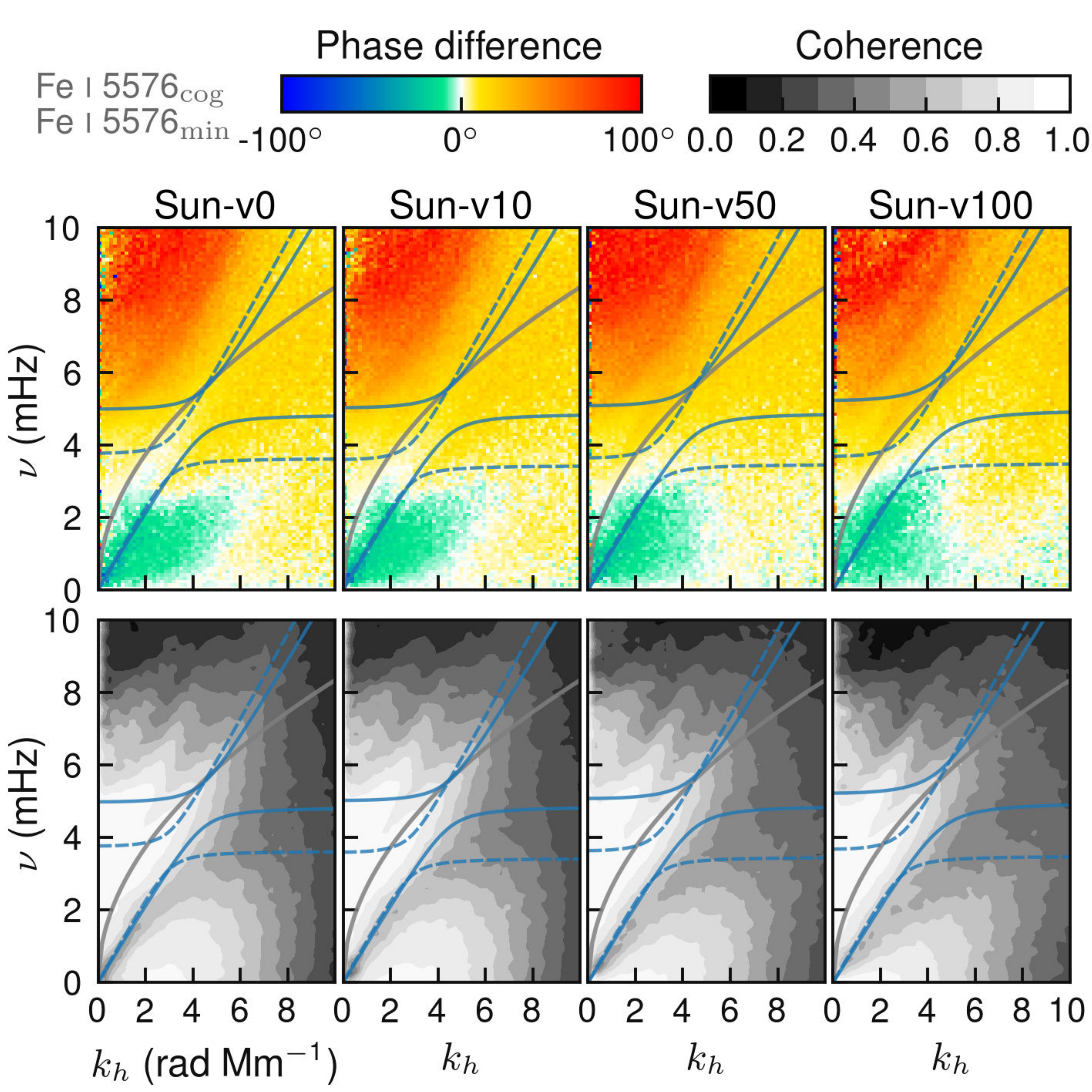}
        \caption{Phase difference (colour) and coherence (greyscale) spectra 
                for the {\ion{Fe}{I}~$\lambda$~5576~\AA{}} line measured with the 
                velocities derived using the COG and line-minima shifts. The 
                grey curve marks the dispersion relation of the surface gravity 
                waves. The blue curves mark the propagation boundary of the 
                acoustic-gravity wave corresponding to the approximate formation 
                heights where the COG (dashed) and line-minima (solid) velocities 
                are measured.}
        \label{fig:ph_ch_spectra_single_5576}
\end{figure}

In Fig.~\ref{fig:ph_ch_spectra_single_5576}, we show the phase and
coherence spectra for the {\ion{Fe}{I}~$\lambda$~5576~\AA{}} line. The
separation between the peak of $RF_{v}$ for the COG and line-minima is
much less in this case ({290--300}~km) compared to that of
the {\ion{Fe}{I}~$\lambda$~5434~\AA{}} line. As a consequence of this, the
coherence is stronger for this line. Additionally, the fact that
{\ion{Fe}{I}~$\lambda$~5576~\AA{}} is less temperature sensitive adds to
the stronger coherence between the layers, as the line-forming layers
are less corrugated. The influence of magnetic field on the phase
difference spectra is barely discernible except for the change in sign
in the IGW region for wave numbers in the range of {4 -- 6~rad~Mm$^{-1}$}
for frequencies above $\nu=$1 mHz (clearly seen when comparing the yellow area in the IGW
region in Sun-v100 to the green area in the Sun-v0 case).

\begin{figure*}[ht!]
    \centering
        \includegraphics[width=.97\columnwidth]{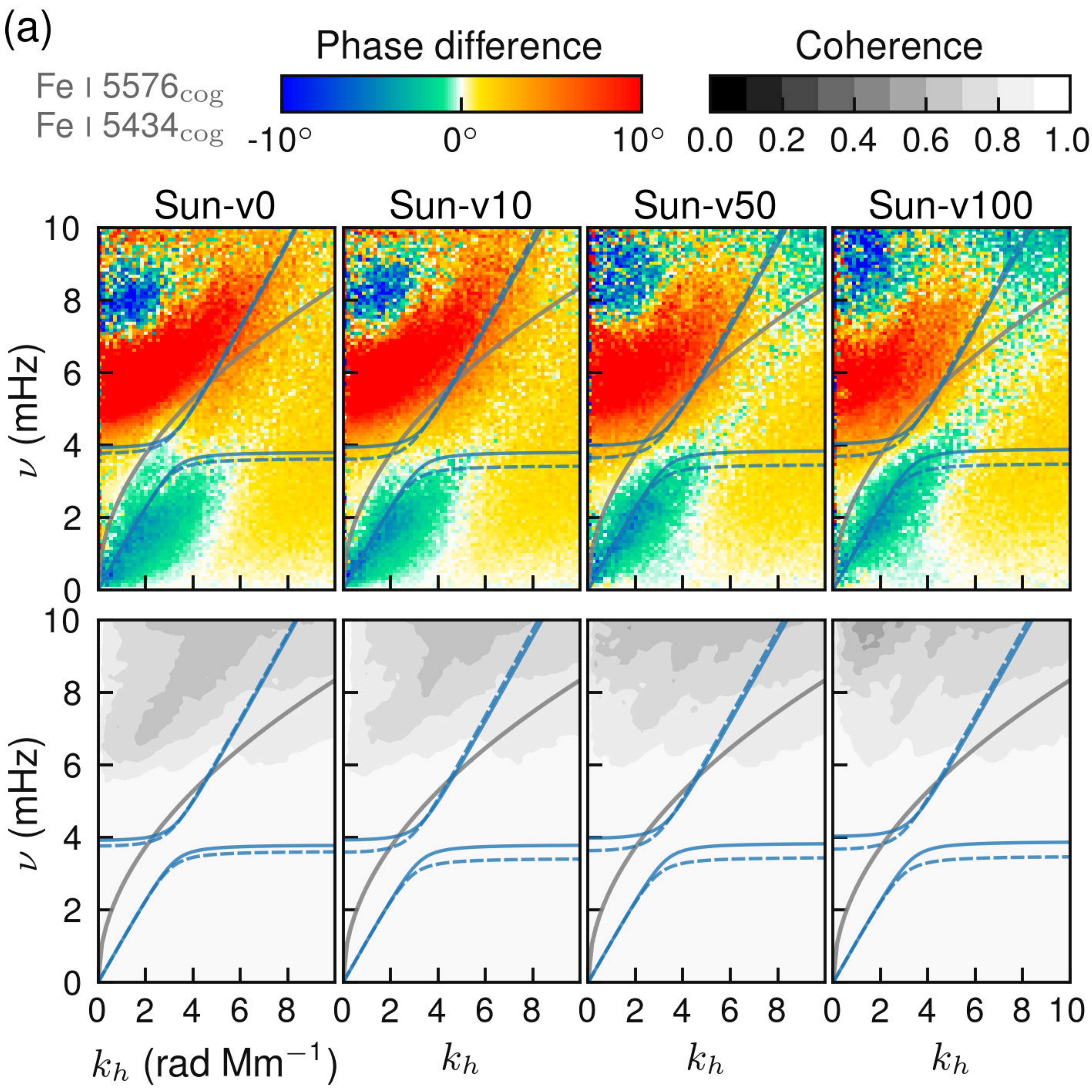}\hskip4ex\includegraphics[width=.97\columnwidth]{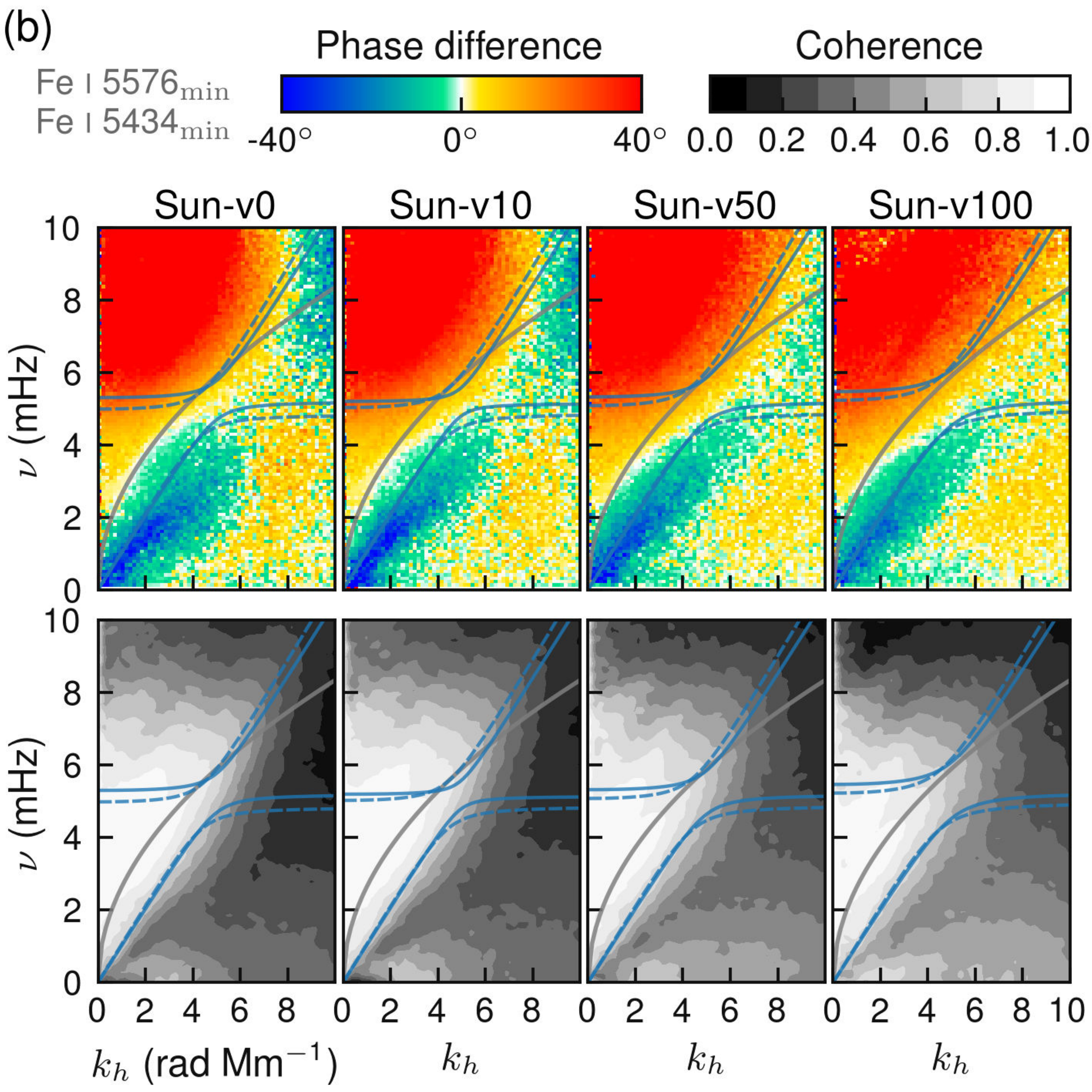}
        \caption{Phase difference (colour) and coherence (greyscale) spectra 
                estimated from the (a) velocities measured using the COG shift 
                of {\ion{Fe}{I}~$\lambda$~5434~\AA{}} and
                {\ion{Fe}{I}~$\lambda$~5576~\AA{}} and (b) velocities measured 
                using the line-minima shift of {\ion{Fe}{I}~$\lambda$~5434~\AA{}} 
                and {\ion{Fe}{I}~$\lambda$~5576~\AA{}}. The grey curve marks 
                the dispersion relation of the surface gravity waves. The blue 
                curves mark the propagation boundary of the acoustic-gravity 
                wave corresponding to the approximate formation heights where 
                the COG (dashed) and line-minima (solid) velocities are measured.}
        \label{fig:ph_ch_spectra_two}
\end{figure*}

We now turn to the phase and coherence spectra as obtained from looking
at velocity signals from two separate spectral lines. In
Fig.~\ref{fig:ph_ch_spectra_two} we show the phase difference between
the {\ion{Fe}{I}~$\lambda$~5576~\AA{}}  and
{\ion{Fe}{I}~$\lambda$~5434~\AA{}} lines using both the COG method (left)
and line-minima (right), separately. We see that the coherency has
increased as a result of probing surfaces separated by a shorter
distance, for instance 30~km for the COG method and 280--310~km for the
line-minima method. This shows that the phase spectra are more reliable
when two different lines are considered that are separated by a rather
close formation height, provided the phase difference is still
measurable. We see that for the line-minima, the phase spectra show a
hint of the effect of magnetic field (the yellow area in the IGW with
the corresponding coherence plot showing stronger coherence in the same
area). This is due to the fact that the line-minima samples the velocity
field at higher layers, where the presence of magnetic fields may lead
to the reflection of these waves resulting in positive phase difference.

\subsection{Dependence of phase difference on vertical travel distance}
\label{ss:phase_height_variation}
The change in the phase spectra for different pairs of heights motivates
us to look at how they vary as a function of vertical distance between
the measurement heights. The dependence of the phase lag and coherence
with distance between the two velocity measurements give us a sense of
the height separation in the atmosphere for which the phase difference
can be reliably measured. To examine this further, we choose a layer as a reference
layer and compute $k_{h}-\omega$ phase difference spectra for different
atmospheric layers relative to this reference layer. We can then pick a
representative $k_{h}$ and $\omega$ that fall within the region where
the bulk of the IGWs emission occurs and show the phase difference as a
function of the travel distance between the reference layer and any
height in the atmosphere.

\begin{figure}[h!]
        \centering
        \includegraphics[width=1.\columnwidth]{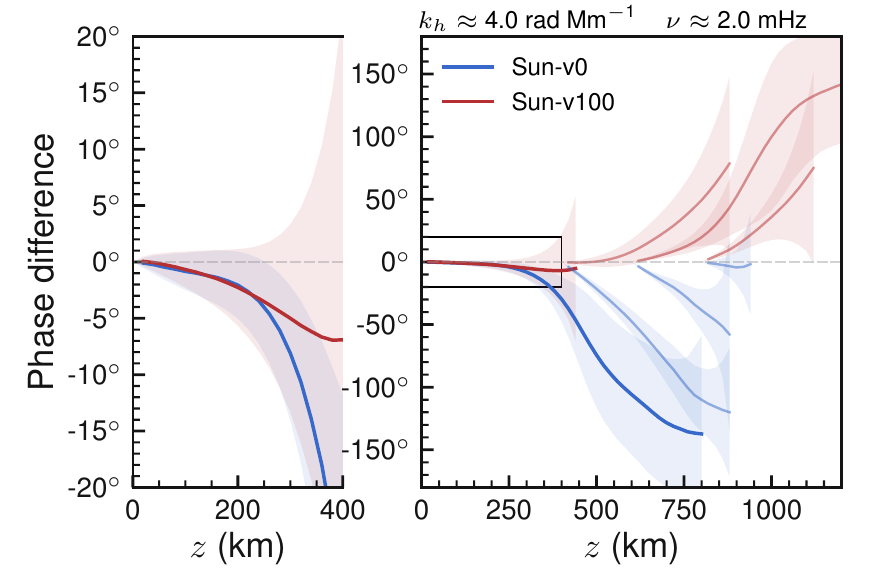}
        \caption{Phase difference between the reference layer, $z$=0, and 
                layers of constant geometrical scale for a given $k_{h}$ and 
                $\omega$ (thick solid red and blue curves). The left plot is 
                a rescaled part of the small region in the right plot. The 
                thin solid lines show three other reference layers 
                ($z=$~400,600,800~km.). The 90\% confidence bounds for the 
                phase difference estimate is represented by the shaded area.}
        \label{fig:phase_travel_height_z}
\end{figure}

In Fig.~\ref{fig:phase_travel_height_z} we show the phase difference
as a function of travel distance relative to the reference height of
$z=0$~km (thick solid curves) for the non-magnetic model (blue: Sun-v0)
and the magnetic model (red: Sun-v100). The phase difference shown here
is computed with the velocity field obtained directly from the
simulation. The 90\% confidence bounds for the phase difference
estimates, according to Sect.~\ref{ss:conf_interval}, are represented by
the shaded area of the corresponding colour. Also shown as thin solid curves
are the phase differences relative to a selection of other reference
layers, $z=400,600,800$~km, along with the corresponding 90\% confidence
bounds. The dependence of the phase propagation on the magnetic field
can be clearly seen in the sign of the phase difference. Both the
non-magnetic and magnetic models show negative phase difference up to
around 300~km, which means that the waves are propagating in the upward
direction. Beyond the vertical travel distance of 400~km for the
$z=0$~km reference layer, the phase measurement become unreliable for
the magnetic case. For the non-magnetic model, the phase difference can
be estimated even beyond a vertical travel distance of 750~km. Comparing
the phase difference with respect to the other reference layers, this
separation distance becomes shorter as we move up in the atmosphere. The
sign of the phase difference in the case of the non-magnetic model
remains negative (upward-propagating IGW), relative to any reference
layer that is considered. On the other hand, for the magnetic models, the
phase propagation relative to the reference height above $z=400$~km is always positive, suggesting the downward propagation of waves in the
presence of magnetic field.

\subsection{Dependence of coherence on vertical travel distance}
Despite the $v$--$v$ phase difference spectra suggesting signatures of
IGW propagation, a definite validation is only possible by looking at
the coherence spectra to help us in judging whether the measured phase
differences are acceptable. The coherence spectra describe the quality
and the reliability of the phase spectra, giving us a measure of the correlation of the two vertical velocity signals. Looking at
Figs.~\ref{fig:ph_ch_spectra_single_5434}-\ref{fig:ph_ch_spectra_two},
the coherence spectra tends to zero in specific regions of the
$k_{h}-\omega$ space, rendering any interpretation of the phase
difference in that Fourier region unreliable. It is clear that the
coherence depends on the height separation between the velocity
measurements, mainly because the waves have to travel a long distance
and they tend to loose their coherency with distance. Another reason
could be that the surfaces of constant optical depth on which the
velocities are measured are very irregular relative to constant phase
surfaces of the wave. Evidence in favour of this hypothesis can be seen when
comparing the coherence spectra of \ion{Fe}{I}~$\lambda$~5434~\AA{}
(Fig.~\ref{fig:ph_ch_spectra_single_5434}) and
\ion{Fe}{I}~$\lambda$~5576~\AA{}
(Fig.~\ref{fig:ph_ch_spectra_single_5576}) which show a strong coherence
for the temperature insensitive
\ion{Fe}{I}~$\lambda$~5576~\AA{} line.

An additional factor that may influence the coherence is the
angle of propagation with respect to the vertical for a wave of a given
frequency, which is strictly governed by the local
Brunt-V\"{a}is\"{a}l\"{a} frequency ($N$). A wave launched at a specific
frequency, in the absence of any non-linear interaction will eventually
follow a curved trajectory if the local $N$ changes with height.
Therefore, the coherency of the waves propagating between two heights
for a given Fourier frequency may locally change over the field of view.
Although the $N$ in our model atmospheres change with height as
described in
\citetalias{2017ApJ...835..148V}, 
we note that it does not differ between the models of different average
field strength, and therefore the difference in phase trajectory between
waves in the models below $\beta=1$ should be negligible. Only in the
higher layers does the effect of magnetic field appear to be strong. Questions that remain refer to how the magnetic field affects the coherence, or whether or not
the coherence changes if magnetic fields are present.

\begin{figure}[h!]
        \centering
        \includegraphics[width=1.\columnwidth]{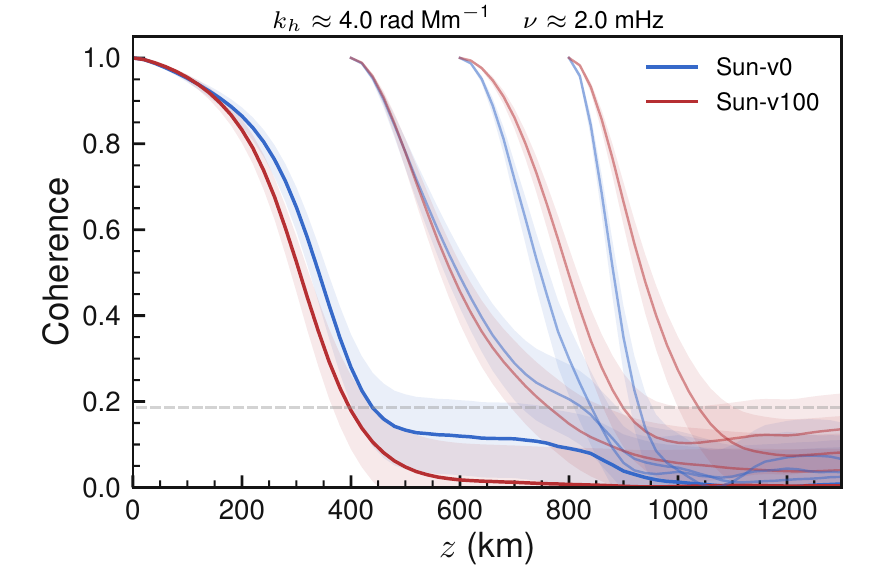}
        \caption{Coherence between the $z$=0 layer and layers of constant 
                geometrical height for a given $k_{h}$ and $\omega$. The 
                zero-coherence threshold at a significance level of 0.05 is 
                marked by the dashed line. The 90\% confidence bounds for the 
                coherence are represented by the shaded area.}
        \label{fig:coherence_height_z}
\end{figure}

Similar to the phase difference study carried out in the previous
section, we now look at the coherence as a function of the vertical
travel distance. Figure~\ref{fig:coherence_height_z} shows the coherence
between a selection of reference heights ($z$ = 0, 400, 600, 800~km) and
the layers above. Also shown are the 90\% confidence bounds for the
coherence estimate shown as background shading. The zero-coherence threshold as
described in Sect.~\ref{ss:zero_coherence_threshold} for the specific
$k_{h}$ is marked by the dashed grey line. Here again we look at
$k_{h}=4$~rad~Mm$^{-1}$ and $\nu=2$~mHz, which falls well within the IGW
excitation region.  We see that the coherency drops slowly for the lower
layers but sharply for the layers above. The height difference
corresponds to around 400~km, which is above the significance level,  for all the
models when looking at the coherency between $z$=0~km and layers above.
However, higher up, the difference corresponds around 100~km for the non-magnetic model and 200~km
for our strongest magnetic model. This analysis shows us that when
computing the phase spectra, it is better to choose spectral lines that
form very close to each other in the atmosphere; especially  in the
case of higher layers.

\subsection {Energy flux}
\begin{figure*}[h!]
        \centering
        \includegraphics[width=0.97\columnwidth]{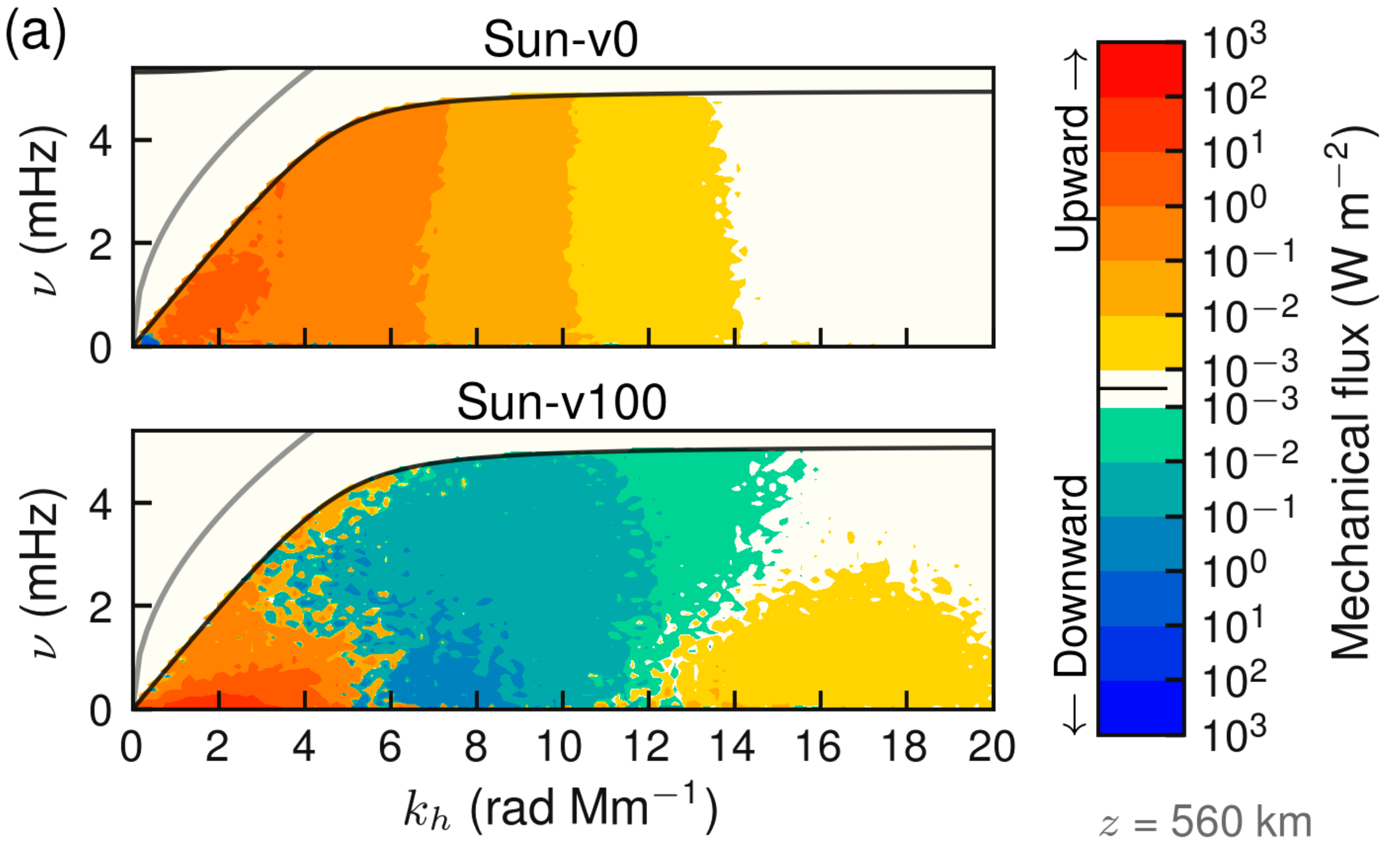}\hskip2ex\includegraphics[width=0.97\columnwidth]{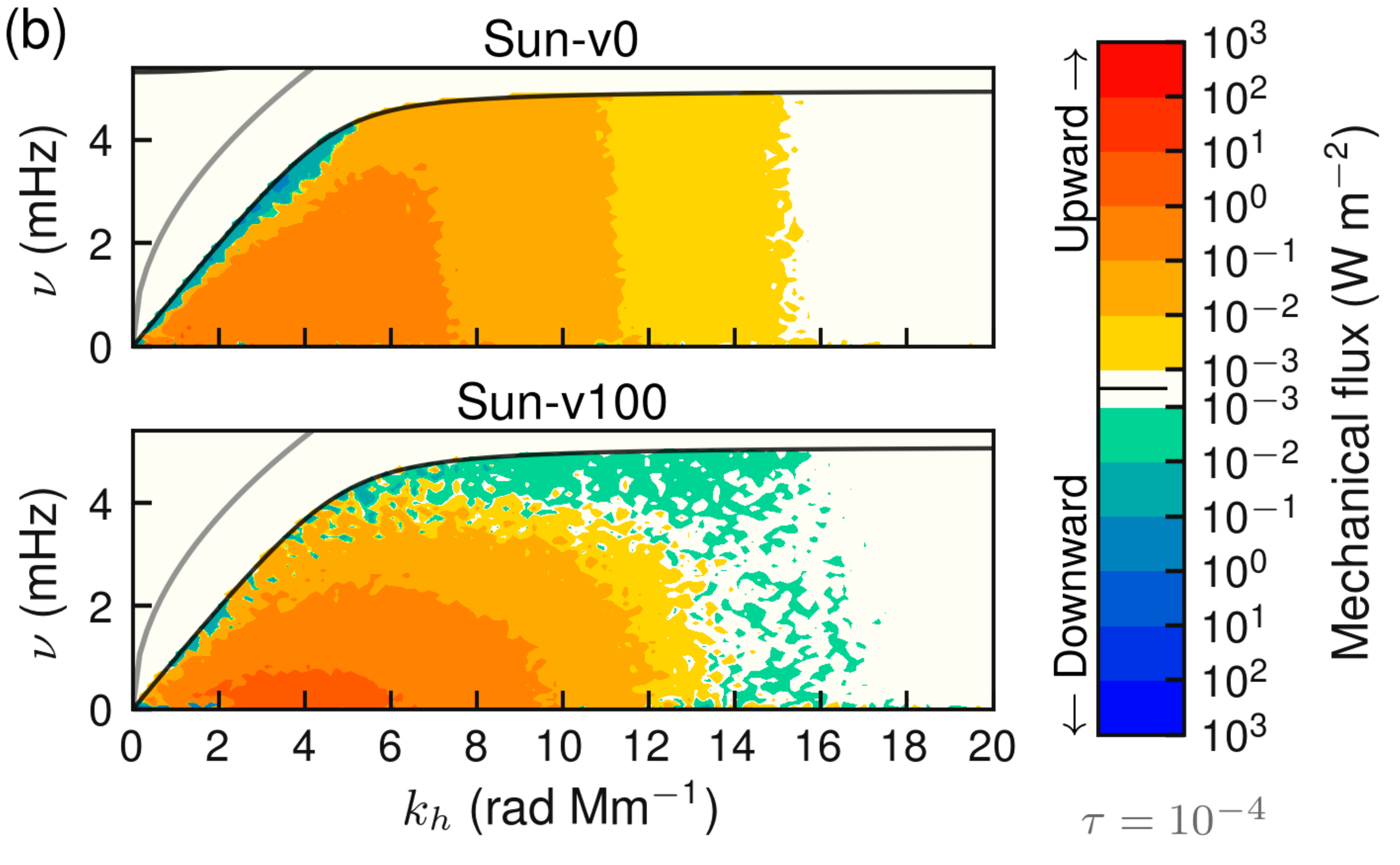}\\
        \includegraphics[width=0.97\columnwidth]{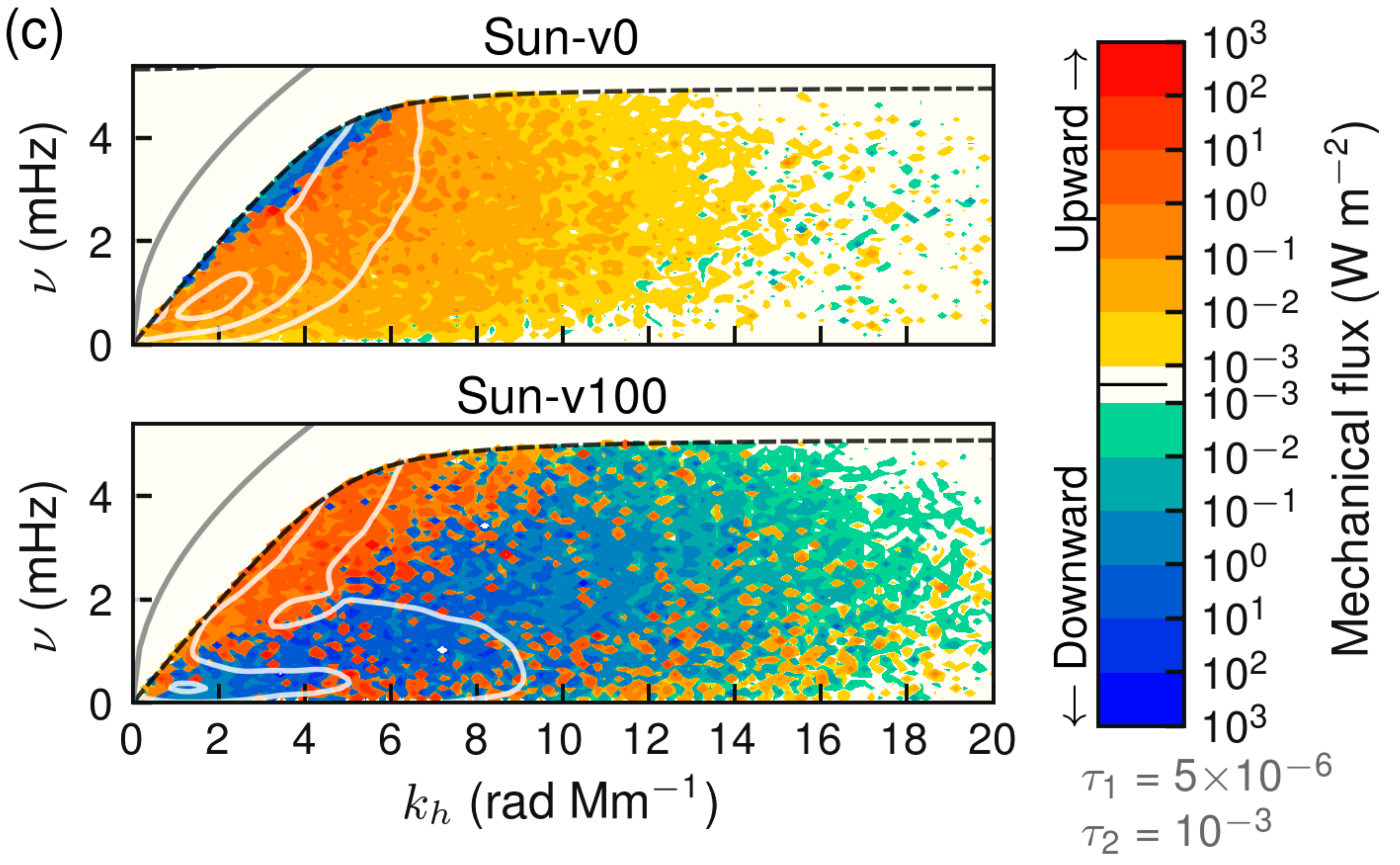}\hskip2ex\includegraphics[width=0.97\columnwidth]{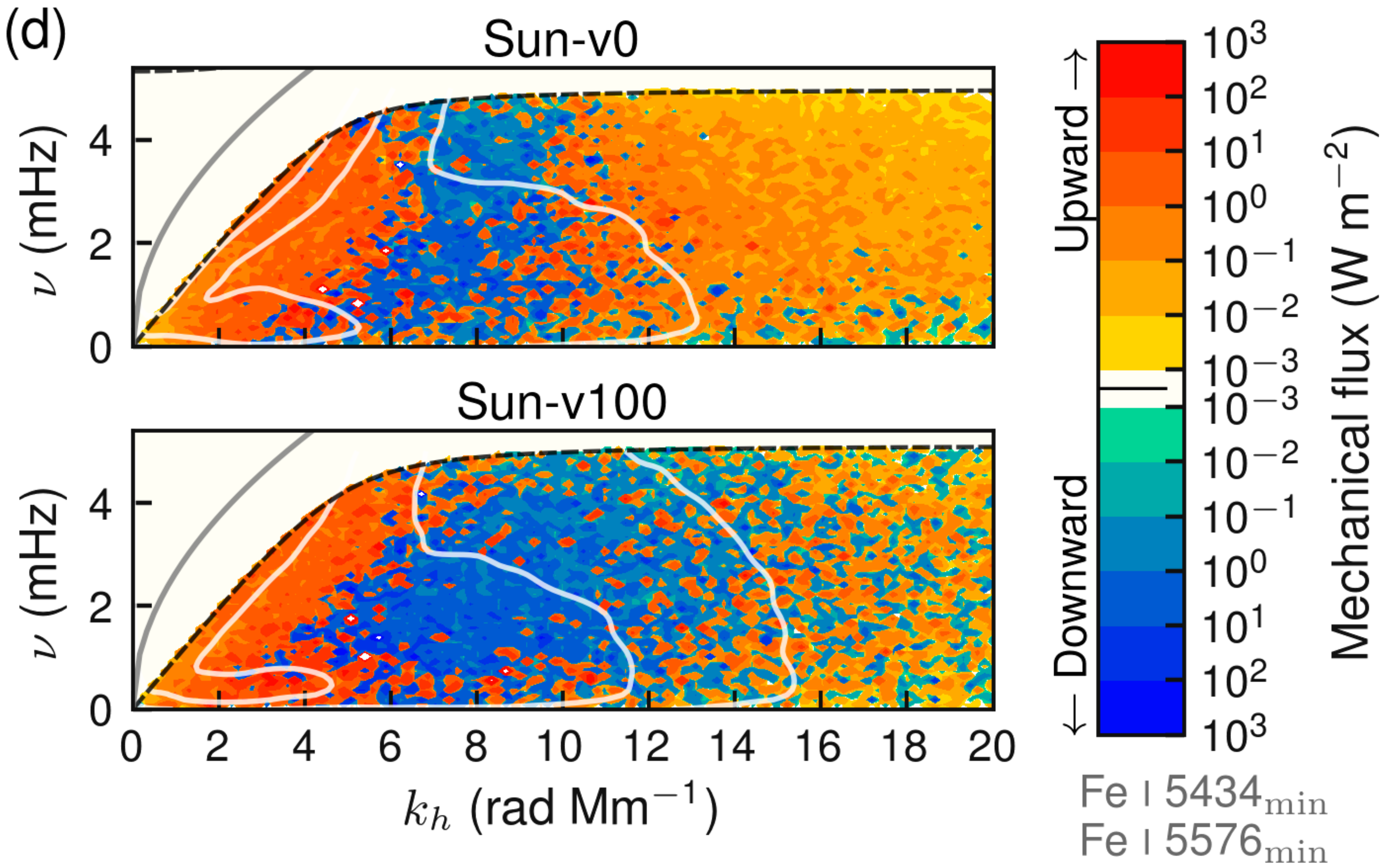}
        \caption{Vertical mechanical flux spectra at (a) a height of $z=560$~km 
                and at (b) the Rosseland mean optical depth layer of 
                $\tau_{R}=1\times$10$^{-4}$. Also shown are the vertical energy flux spectra 
                estimated from phase difference measurements of (c) velocities at 
                two layers of  constant optical depth:  $\tau_{R}=5\times10^{-6}$ and 
                $\tau_{R}=1\times10^{-3}$ and from (d) the line-minima shifts of 
                the {\ion{Fe}{I}~$\lambda$~5434~\AA{}} and 
                {\ion{Fe}{I}~$\lambda$~5576~\AA{}} lines. The plot is masked to 
                show only the IGW region. The grey curve marks the dispersion 
                relation of the surface gravity waves. The black dashed curves 
                mark the propagation boundary of acoustic-gravity waves 
                corresponding to approximately the middle region of the two 
                layers.}
        \label{fig:energy_flux_tau_lines}
\end{figure*}

A first-order approximation to the flux of the total energy density
associated with IGWs can be directly determined from the simulation.
This `linear' wave energy flux is given by the perturbed
pressure–velocity ($\Delta p-v$) co-spectrum (Eq.~7 of
\citetalias{2017ApJ...835..148V})
in the IGW region. In Fig.~\ref{fig:energy_flux_tau_lines}(a), we show
the vertical flux of the active mechanical energy density for the
non-magnetic (Sun-v0) and magnetic (Sun-v100) models at a height of
$z=$560~km. The main difference between the two is the presence of a
downward flux (green-blue) in the magnetic model within the IGW branch.
We obtain a total positive flux of $\sim$830 {W~m$^{-2}$} for the Sun-v0
and $\sim$1475 {W~m$^{-2}$} for the Sun-v100 model in the IGW region for
k$_{h}<10$~rad~Mm$^{-1}$. The energy flux for a layer of constant
optical depth, as shown in Fig.~\ref{fig:energy_flux_tau_lines}(b) for
$\tau_{R}=1\times$10$^{-4}$, gives $\sim$320~{W~m$^{-2}$} for the Sun-v0
and $\sim$617~{W~m$^{-2}$} for the Sun-v100 model. The main discrepancy
between the two flux estimates is due to the fact that the pressure
perturbations are weaker in layers of constant optical depth compared to
the fluctuations of pressure at a given geometrical height.
        
We can now compare this with the estimates obtained using
phase-difference spectra, where the energy flux for IGWs is determined
from the mean energy density and the group velocity, as described in
Sect.~\ref{ss:analysis:energy_flux}. In
Fig.~\ref{fig:energy_flux_tau_lines}(c), we show the energy flux as
determined from the phase difference between two layers of constant
optical depth, $\tau_1$=$5\times10^{-6}$ and  $\tau_2$=$1\times10^{-3}$
for the non-magnetic (Sun-v0) and magnetic (Sun-v100) models. These
velocities are directly obtained from the simulation box. The difference
in the $\tau_{\rm R}$ level corresponds to a height difference of
$\Delta z=280-290$~km for the non-magnetic and the magnetic models. The
white contours mark selected coherence values, namely 0.2, 0.4, and 0.6.
Comparing the non-magnetic and the magnetic models, we do see the change
in the direction of energy transport, corresponding to upward in the
non-magnetic case and a mixture of upward and downward in the magnetic
case. We also note that the location of strong $v$--$v$ coherence in the
$k_{h}$-$\omega$ plane is different in both cases, as marked by the
profile of the coherence contours. However, comparing this with the flux
spectra shown in {Figs.~\ref{fig:energy_flux_tau_lines}(a) and (b)}, we
see significant differences, particularly in the case of the magnetic
model. The total positive flux that we obtain with the phase difference
between two layers of constant optical is $\sim$1200~{W~m$^{-2}$} and
$\sim$8400~{W~m$^{-2}$} for Sun-v0 and Sun-v10, respectively. The
magnetic model shows a significant difference in the total flux between
the phase-difference analysis and direct estimation.
        
Finally, Fig.~\ref{fig:energy_flux_tau_lines}(d) shows the energy flux
as determined from the line core Doppler velocities of
{\ion{Fe}{I}~$\lambda$~5434~\AA{}} and
{\ion{Fe}{I}~$\lambda$~5576~\AA{}}. The propagation boundaries in both
the plots correspond to a height of $z\approx$550~km. The mean density
and the mean Brunt-V\"{a}is\"{a}l\"{a} frequency are taken from this
layer for the estimation of the energy flux. A height separation
($\Delta z$) of 310~km for the non-magnetic model and 280~km for the
magnetic model is assumed. The total positive flux we obtain for the IGW
using the spectral line is an order of magnitude more than what we
estimate from simulations using the same method. For the Sun-v0 model we
obtain a total positive flux of $\sim$38200~{W~m$^{-2}$} and for the
Sun-v100 model we get $\sim$90900~{W~m$^{-2}$}. The velocities from the
simulation at  layers of constant optical depth are comparable to the
Doppler velocities that we derive from the spectral line. However, we
see that the phase differences in the simulation are significantly
different from the phase differences obtained using the spectral lines.
They are considerably smaller, with a value often close to zero, leading to very high
phase speed and therefore very high group speed estimates. Energy flux
estimates based on this method may therefore be significantly
overestimated.

To summarise, we note that the energy flux from the simulation and that
obtained using the spectral lines for the non-magnetic model are
significantly different in terms of both the region where the strongest
flux lies and its magnitude. The magnetic models show a larger
difference compared to the non-magnetic model. The actual reason for the
differences in the energy flux spectra measured with the different
methods is still unclear. However, we would like to point out that
relying solely on the energy estimates from the phase difference
measurement may lead to unrepresentative results. In a realistic atmosphere, surfaces of
constant optical depth show large variations. In addition to this, the
velocity fields are required at two separate layers in order to
determine the energy flux. The velocity field is likely to be smeared
out resulting in inaccurate determination of the phase difference, and
consequently the energy flux.

\subsection{Comparing wave spectra from simulations with synthetic observations}

In 
\citetalias{2017ApJ...835..148V} 
and 
\citetalias{2019ApJ...872..166V}, 
where we studied the wave spectra directly in the simulation domain, we
clearly see the influence of the magnetic field on the propagation of
IGWs with height. However, as this work shows, when it comes to
synthetic observation, the possible signatures of the magnetic fields on
the IGW propagation are not apparent. Although we clearly see signatures
of IGWs in the synthetic Doppler velocities, it is hard to discern the
effect of magnetic fields on the propagation of these waves unless
velocity measurements from very near-lying layers are used.

The observational detection of IGWs are further challenged by the
problem of coherence. In the models considered here, a reliable IGW
spectrum can be measured in the near surface layers with a height
difference of up to 400~km, while in the chromosphere this drastically
reduces to less than 100~km. Given the width of formation of the line,
it is quite challenging to obtain a reliable phase difference
measurement in the low-chromospheric regions. Also, the more temperature
sensitive line,  \ion{Fe}{I}~5434 \AA,~shows stronger incoherence
than the  \ion{Fe}{I}~5576~\AA~line.

The differences between the simulation and the synthetic
observations are mostly attributable to the fact that the waves are probed at different layers.
While the geometrical height is unchanged by the passage of the waves,
the optical depth scale varies and may be modified by the waves, making
the coherence weak and the corresponding phase difference analysis
difficult. Additionally, in the geometrical height scale we have
measurements from precise heights, whereas in the synthetic observation case our
velocity measurements are taken over a broader region and are often
smeared out. This also adversely affects the estimation of the
energy flux from spectral lines.

\subsection{Comparing wave spectra from synthetic with real observations}
The measurements from the synthetic Dopplergram facilitates a direct
comparison to observations. Our results are consistent with the
observations by
\cite{2008ApJ...681L.125S} 
\&  
\cite{2014SoPh..289.3457N} 
in terms of the qualitative properties of the phase-difference. In order
to directly compare our results with observations, we compare our
results with those of
\cite{2011A&A...532A.111K}.\linebreak
They
looked at the \ion{Fe}{I}~$\lambda\lambda$~5576~\AA~and~5434 \AA~lines,
which were scanned quasi-simultaneously. These latter authors estimated the power,
phase, and coherence spectra from the line minimum velocities for the
two lines. They report a minimum phase difference of approximately
$-40\degr$ for $k_{h}\approx4$~rad Mm$^{-1}$ and $\nu\approx2.5$~mHz,
which is consistent with our results as evident from
Fig.~\ref{fig:ph_ch_spectra_two}(b). The authors report that for
$k_{h}\ge$13~rad Mm$^{-1}$ the phases become random, which is also seen
in our synthetic data (not shown in Fig. 9(b)). In terms of coherence,
they obtain a maximum of 0.85 in the IGW regime, which is slightly
higher than the coherence we see in our simulated models and the
spectral lines. Our study provides an indication that the simulations
reproduce the observations and confirm the existence of IGWs.
\subsection{Future observational strategy}
Our study of the two neutral Fe lines shows that for the analysis of lines with a
high-excitation potential, magnetically non-sensitive lines may be
better for studying IGWs. Some lines possibly well-suited for
studying the IGWs, 
apart from the two lines studied here, are \ion{Ni}{I}~4912,
\ion{Fe}{I}~5691, \ion{Fe}{I}~7090, and \ion{Fe}{II}~722.45 
\citep{2005A&A...439..687C}. We suggest
future observations to look at either a single weak line and measure the
phase and coherence from the different Doppler velocities, or to look at
two spectral lines that are in approximately the same region in the
atmosphere. Ongoing work with predominantly horizontal fields shows
upward propagation for IGWs (downward phases) even in low
plasma-$\beta$, indicating that these waves may still be important for
chromospheric layers. We infer from our analysis that the photospheric
detection of IGWs is relatively straightforward, but observing waves in the higher
layers might be hindered by the availability of chromospheric lines that
have a narrow line-of-formation profile. Progress may come from exploring other techniques that provide more localised
measurement of the solar velocity field, like relying on a combination
of multiple lines
\citep{2002A&A...395L..51W}. 
Furthermore, any analysis should certainly consider NLTE as evident from
the difference in the $RF_{v}$ that we estimate using LTE.

\section{Conclusions}\label{s:conclusion}
Here, we present a study of internal gravity waves detected in
synthetic observations that cover a range of quiet solar
magneto-atmospheric conditions. A selection of models of varying average
magnetic flux density were simulated with the state-of-the-art
radiative-MHD code {CO$^{\rm 5}$BOLD}. Using these time-evolving 3D
dynamic models as input, we synthesised spectral lines at individual
snapshots to look at signatures of waves in the intensity and derived
Doppler velocity maps. We compared the wave spectra in the simulated
physical volume with that obtained from the synthetic observations. In
this study, we relied on the different estimates of the Doppler velocities
to infer the 2D velocity field at various layers in the solar atmosphere
in order to carry out the phase and coherence analyses. One of our main
aims was to investigate the feasibility of detecting signatures of IGWs
in the emergent wave spectra. Furthermore, we were motivated to study
the influence of magnetic field on the waves and its manifestation in
the diagnostic spectra. In addition to gaining an  understanding of how to detect IGWs in real
observations, we also aimed to examine the method for observational estimation
of the energy flux of these waves, an aspect which is important in the
context of their contribution to the total wave energy budget of the
solar atmosphere.

The time series of the Doppler velocities estimated from two
non-magnetic neutral iron lines \ion{Fe}{I}~$\lambda\lambda$ 5434
\AA~and 5576 \AA~ were used for the study of waves. We carried out
Fourier analyses to derive the power, phase, and coherence spectra in the
$k_{h}-\omega$ dispersion relation diagram to separate IGWs from
acoustic waves. The characteristic properties of the phases in the
dispersion relation diagram clearly show us that the waves are indeed
observable in velocity maps obtained from observations. However, the
coherency varies strongly within different regions of the IGW domain
rendering any phase analysis in those regions unreliable. Furthermore,
any detection of the effect of magnetic field on these waves is
complicated by the properties of the selected lines and their formation
height. Our analysis shows that unlike the clear signature of IGW
propagation seen in simulations, the observation signatures do not
allow for a straightforward interpretation on the effect of magnetic
field on the propagation of these waves. By comparing energy flux directly
obtained from the simulation with that determined from the spectral
lines, we come to the conclusion that the specific choice of spectral
lines plays an important role;  the types of lines needed for such an analysis tend to smear out the velocity
field depending on the width of the line-formation region and their
spatial separation, resulting in significantly smaller phase differences
which can lead to overestimation of the
energy flux of IGWs by an order of magnitude. Our understanding is that lines that are
temperature insensitive and non-magnetic with narrow line-formation ranges
may be preferable for detecting and studying internal waves
in the solar atmosphere.

\begin{acknowledgements}
          GV would like to thank B. Fleck, O. Steiner, J.M. Borrero, 
          N. Bello Gonz\'{a}lez, A. Pastor Yabar, and J. Bruls for the 
          helpful discussions. We thank the referee for his detailed 
          comments and suggestions that significantly improved the 
          presentation of the paper. This work was supported by the
      {Deut\-sche For\-schungs\-ge\-mein\-schaft, DFG\/} grant 
      RO 3010/3-1. 
\end{acknowledgements}

\bibliographystyle{aa} 
\bibliography{} 

\end{document}